\documentclass[a4paper,fleqn]{cas-sc}

\usepackage[numbers]{natbib} 
\def\tsc#1{\csdef{#1}{\textsc{\lowercase{#1}}\xspace}}
\tsc{WGM}
\tsc{QE}

\usepackage[english]{babel}
\usepackage{siunitx}
\usepackage{cleveref}

\usepackage{amsmath, esint}
\usepackage{graphicx}
\usepackage{subcaption}
\usepackage{amssymb}
\usepackage{listings}
\usepackage{hyperref}
\usepackage{blindtext}
\usepackage{lineno}
\usepackage{array}
\usepackage{multirow}

\shorttitle{MeVPrtl: An Event Generator for Dark Sector Particles in the SBN Program} 
\title[mode = title]{MeVPrtl: An Event Generator for Dark Sector Particles in the Short-Baseline Neutrino Program}

\newcommand{\ANL}{Argonne National Laboratory, Lemont, IL 60439, USA}
\newcommand{\Bern}{Universit\"{a}t Bern, Bern CH-3012, Switzerland}
\newcommand{\BNL}{Brookhaven National Laboratory, Upton, NY 11973, USA}
\newcommand{\UCSB}{University of California, Santa Barbara CA, 93106, USA}
\newcommand{\Campinas}{Universidade Estadual de Campinas, Campinas, SP 13083-970, Brazil}
\newcommand{\CTI}{Center for Information Technology Renato Archer, Campinas, SP 13069-901, Brazil}
\newcommand{\Chicago}{University of Chicago, Chicago, IL 60637, USA}
\newcommand{\CIEMAT}{CIEMAT, Centro de Investigaciones Energ\'{e}ticas, Medioambientales y Tecnol\'{o}gicas, Madrid E-28040, Spain}
\newcommand{\CSU}{Colorado State University, Fort Collins, CO 80523, USA}
\newcommand{\Columbia}{Columbia University, New York, NY 10027, USA}
\newcommand{\Edinburgh}{University of Edinburgh, Edinburgh EH9 3FD, United Kingdom}
\newcommand{\ABC}{Universidade Federal do ABC, Santo Andr\'{e}, SP 09210-580, Brazil}
\newcommand{\Alfenas}{Universidade Federal de Alfenas, Po\c{c}os de Caldas, MG 37715-400, Brazil}
\newcommand{\FNAL}{Fermi National Accelerator Laboratory, Batavia, IL 60510, USA}
\newcommand{\Florida}{University of Florida, Gainesville, FL 32611, USA}
\newcommand{\Granada}{Universidad de Granada, Granada E-18071, Spain}
\newcommand{\IIT}{Illinois Institute of Technology, Chicago, IL 60616, USA}
\newcommand{\Imperial}{Imperial College London, London SW7 2AZ, United Kingdom}
\newcommand{\IIS}{Centre for High Energy Physics, Indian Institute of Science, Bangalore 560012, India}
\newcommand{\SaoJose}{Instituto Tecnológico de Aeronáutica, São José dos Campos, SP 12228-900, Brazil}
\newcommand{\Kansas}{University of Kansas, Lawrence, KS 66045, USA}
\newcommand{\Lancaster}{Lancaster University, Lancaster LA1 4YW, United Kingdom}
\newcommand{\Liverpool}{University of Liverpool, Liverpool L69 7ZE, United Kingdom}
\newcommand{\LANL}{Los Alamos National Laboratory, Los Alamos, NM 87545, USA}
\newcommand{\LSU}{Louisiana State University, Baton Rouge, LA 70803, USA}
\newcommand{\Manchester}{University of Manchester, Manchester M13 9PL, United Kingdom}
\newcommand{\Michigan}{University of Michigan, Ann Arbor, MI 48109, USA}
\newcommand{\Minnesota}{University of Minnesota, Minneapolis, MN 55455, USA}
\newcommand{\Holyoke}{Mount Holyoke College, South Hadley, MA 01075, USA}
\newcommand{\NotreDame}{University of Notre Dame, Notre Dame, IN 46556 USA}
\newcommand{\Oxford}{University of Oxford, Oxford OX1 3RH, United Kingdom}
\newcommand{\Penn}{University of Pennsylvania, Philadelphia, PA 19104, USA}
\newcommand{\Palermo}{Universit\`{a} degli Studi di Palermo, Dipartimento di Fisica e Chimica, I-90123 Palermo, Italy}
\newcommand{\QueenMary}{Queen Mary University of London, London E1 4NS, United Kingdom}
\newcommand{\Rutgers}{Rutgers University, Piscataway, NJ, 08854, USA}
\newcommand{\Sheffield}{University of Sheffield, School of Mathematical and Physical Sciences, Sheffield S3 7RH, United Kingdom}
\newcommand{\Sussex}{University of Sussex, Brighton BN1 9RH, United Kingdom}
\newcommand{\Syracuse}{Syracuse University, Syracuse, NY 13244, USA}
\newcommand{\UTK}{University of Tennessee at Knoxville, TN 37996, USA}
\newcommand{\TAMU}{Texas A\&M University, College Station, TX 77843, USA}
\newcommand{\UTA}{University of Texas at Arlington, TX 76019, USA}
\newcommand{\Tufts}{Tufts University, Medford, MA, 02155, USA}
\newcommand{\UCL}{University College London, London WC1E 6BT, United Kingdom}
\newcommand{\VirginiaTech}{Center for Neutrino Physics, Virginia Tech, Blacksburg, VA 24060, USA}
\newcommand{\Warwick}{University of Warwick, Coventry CV4 7AL, United Kingdom}
\newcommand{\CBPFCENTROBRASILEIRO}{CBPF, Centro Brasileiro de Pesquisas Fisicas, Rio de Janeiro, Brazil}
\newcommand{\CERNEUROPEANORGANIZA}{CERN, European Organization for Nuclear Research 1211 Gen\`eve 23, Switzerland, CERN}
\newcommand{\CENTRODEINVESTIGACIO}{Centro de Investigacion y de Estudios Avanzados del IPN (Cinvestav), Mexico City}
\newcommand{\UNIVERSITYOFHOUSTONH}{University of Houston, Houston, TX 77204, USA}
\newcommand{\INDIANINSTITUTEOFSCI}{Indian Institute of Science, Bengaluru, India}
\newcommand{\INFNSEZIONEDIBOLOGNA}{INFN Sezione di Bologna and University, Bologna, Italy}
\newcommand{\INFNSEZIONEDICATANIA}{INFN Sezione di Catania and University, Catania, Italy}
\newcommand{\INFNSEZIONEDIGENOVAA}{INFN Sezione di Genova and University, Genova, Italy}
\newcommand{\INSTITUTODINEURO}{Istituto di Neuroscienze, CNR, Padova, Italy}
\newcommand{\INFNGSSILAQUILAITALY}{INFN GSSI, L’Aquila, Italy}
\newcommand{\INFNLNGSASSERGIITALY}{INFN LNGS, Assergi,  Italy}
\newcommand{\INFNLNSCATANIAITALY}{INFN LNS, Catania, Italy}
\newcommand{\INFNSEZIONEDIMILANOM}{INFN Sezione di Milano, Milano, Italy}
\newcommand{\INFNSEZIONEDIMILANOB}{INFN Sezione di Milano Bicocca and University, Milano, Italy}
\newcommand{\INFNSEZIONEDINAPOLIN}{INFN Sezione di Napoli, Napoli, Italy}
\newcommand{\INFNSEZIONEDIPADOVAA}{INFN Sezione di Padova and University, Padova, Italy}
\newcommand{\INFNSEZIONEDIPAVIAAN}{INFN Sezione di Pavia and University, Pavia, Italy}
\newcommand{\INFNSEZIONEDIPISAPIS}{INFN Sezione di Pisa, Pisa, Italy}
\newcommand{\IPSIINAF}{IPSI INAF Torino, Italy}
\newcommand{\UNIVERSITYOFPITTSBUR}{University of Pittsburgh, Pittsburgh, PA 15260, USA}
\newcommand{\UNIVERSITYOFROCHESTE}{University of Rochester, Rochester, NY 14627, USA}
\newcommand{\SLACNATIONALACCELERA}{SLAC National Accelerator Laboratory, Menlo Park, CA 94025, USA}
\newcommand{\SOUTHERNMETHODISTUNI}{Southern Methodist University, Dallas, TX 75275, USA}
\newcommand{\YORKUNIVERSITYTORONT}{York University, Toronto, Canada}

\author[41]{P.~Abratenko}
\author[9]{N.~Abrego-Martinez}
\author[12]{R.~Acciarri}
\author[53]{A.~Aduszkiewicz}
\author[63]{F.~Akbar}
\author[26]{D.~Andrade~Aldana}
\author[67]{L.~Aliaga-Soplin}
\author[53]{F.~Abd~Alrahman}
\author[5]{R.~Alvarez-Garrote}
\author[55]{C.~Andreopoulos}

% ...
\address[5]{\CIEMAT}
% ...
\address[9]{\CENTRODEINVESTIGACIO}
% ...
\address[12]{\FNAL}
% ...
\address[26]{\IIT}
% ...
\address[41]{\Tufts}
% ...
\address[53]{\UNIVERSITYOFHOUSTONH}
% ...
\address[55]{\Liverpool}
% ...
\address[63]{\UNIVERSITYOFROCHESTE}
% ...
\address[67]{\UTA}
% ...

\shortauthors{ICARUS and SBND Collaborations for the SBN Program}

\date{\today}

\begin{document}
\begin{abstract}
\noindent\lstinline{MeVPrtl} is a modular event generator of beyond the Standard Model (BSM) physics particles developed for use in the Short-Baseline Neutrino (SBN) Program. A large class of BSM physics models predict that new particles could be produced in the intense Booster Neutrino Beam (BNB) and Neutrinos at the Main Injector (NuMI) beams at Fermilab, travel to the SBN Program detectors, and decay into Standard Model (SM) particles. These new physics models are motivated by dark matter, the neutrino mass scale, and a solution to the strong CP problem. \lstinline{MeVPrtl} provides an interface to implement the overlapping phenomenology of these models, and to connect them
with meson flux inputs and object outputs used by the SBN Program's \lstinline{LArSoft}-based detector simulation. Implementations for the Higgs portal, heavy neutral lepton, and heavy QCD axion models exist within \lstinline{MeVPrtl}. In this paper these implementations and their validation, as well as details of the \lstinline{MeVPrtl} interface, are specified.
\end{abstract} 
\collaboration[ICARUS Collaboration and SBND Collaboration for the SBN Program]
\maketitle

\tableofcontents
\section{Introduction}
\label{sec:intro}

In recent years, neutrino experiments have been shown to be sensitive to numerous dark sector particles (\cite{Batell:2022xau} for an overview). A dark sector particle is a hypothetical new particle that is neutral under the Standard Model (SM) forces. Such particles can be used to probe light dark matter (e.g.\ \cite{BtoS_lightDMsearchViaHiggsPortal}), describe reheating in the early universe (e.g.\ \cite{Scalar_production_channels}), or resolve puzzles unexplained by the SM (e.g.\ \cite{AxionDUNE}). Due to the intense particle beams visible at the Short-Baseline Neutrino (SBN) Program at Fermilab~\cite{MicroBooNE:2015bmn}, and the excellent reconstruction capabilities of liquid argon time-projection chambers (LArTPCs), the SBN Program can have leading sensitivity to certain dark sector models with mediator particle masses in the 100s of $\text{MeV}/c^2$ range. The dark sector particles can be copiously produced in neutrino beams at Fermilab, propagate into the detectors, and decay or otherwise interact to result in visible signals. Further, the signals from such massive particles will arrive at the SBN Program detectors later than those associated with nearly massless neutrinos. In order to further study these models and determine sensitivity, accurate event generation is required. This paper presents the \lstinline{MeVPrtl} event generation code which simulates the chain of dark sector production to decay in the SBN Program detectors, and which is fully integrated into the experiments' analysis software pipeline.

The simplest interactions for the dark sector particles with the SM are portal interactions, in which the dark sector particle mixes with a neutral SM state. Depending on the spin and parity properties, the portal particles fall into a few generic categories: scalars, fermions, vectors, and pseudoscalars. Even if these portal particles decay to the same kinds of visible, SM particles, the production chain for each model and resulting final state kinematics can be quite different, warranting the need for a model-specific event generation tool.

To date, \lstinline{MeVPrtl} simulates production, propagation, and decay in the SBN Program detectors, complete with timing information and probabilistic event weighting, for the following distinct beyond the Standard Model (BSM) models: Higgs portal scalars (HPSs), heavy neutral leptons (HNLs), and heavy QCD axions (which will sometimes be referred to as axion-like particles, or ``ALPs''). The \lstinline{MeVPrtl} interface is designed to be extensible to new model implementation. Beyond the basic portals described here, there are other models that have been proposed for SBN Program and other neutrino experiments with similar phenomenology. Examples of such models include leptophilic axion-like particles \cite{Bertuzzo:2022fcm} and dark photons \cite{DeRomeri:2019kic}. Implementing these models in \lstinline{MeVPrtl} is a possible future avenue of work.

The remainder of this paper is structured as follows. A broad overview of the structure of \lstinline{MeVPrtl} is presented in \cref{sec:tech}.  Each simulation stage is described in detail in \cref{sec:simulation}.  The physics of each of the implemented models is described in detail in \cref{sec:models}.  Finally, future use of this tool is disussed in \cref{sec:discussion}.

\section{\lstinline{MeVPrtl} Overview}
\label{sec:tech}
\begin{figure}[t]
    \centering
    \includegraphics[width=\textwidth]{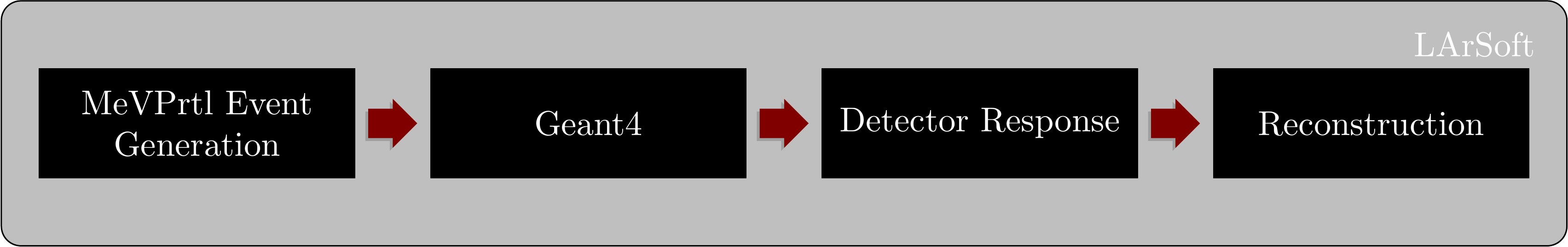}
    \caption{Diagram of how the \lstinline{MeVPrtl} generator fits into SBN Program Monte Carlo production.}   
    \label{fig:mcgen}

    \vspace{1cm}

    \centering
    \includegraphics[width=\textwidth]{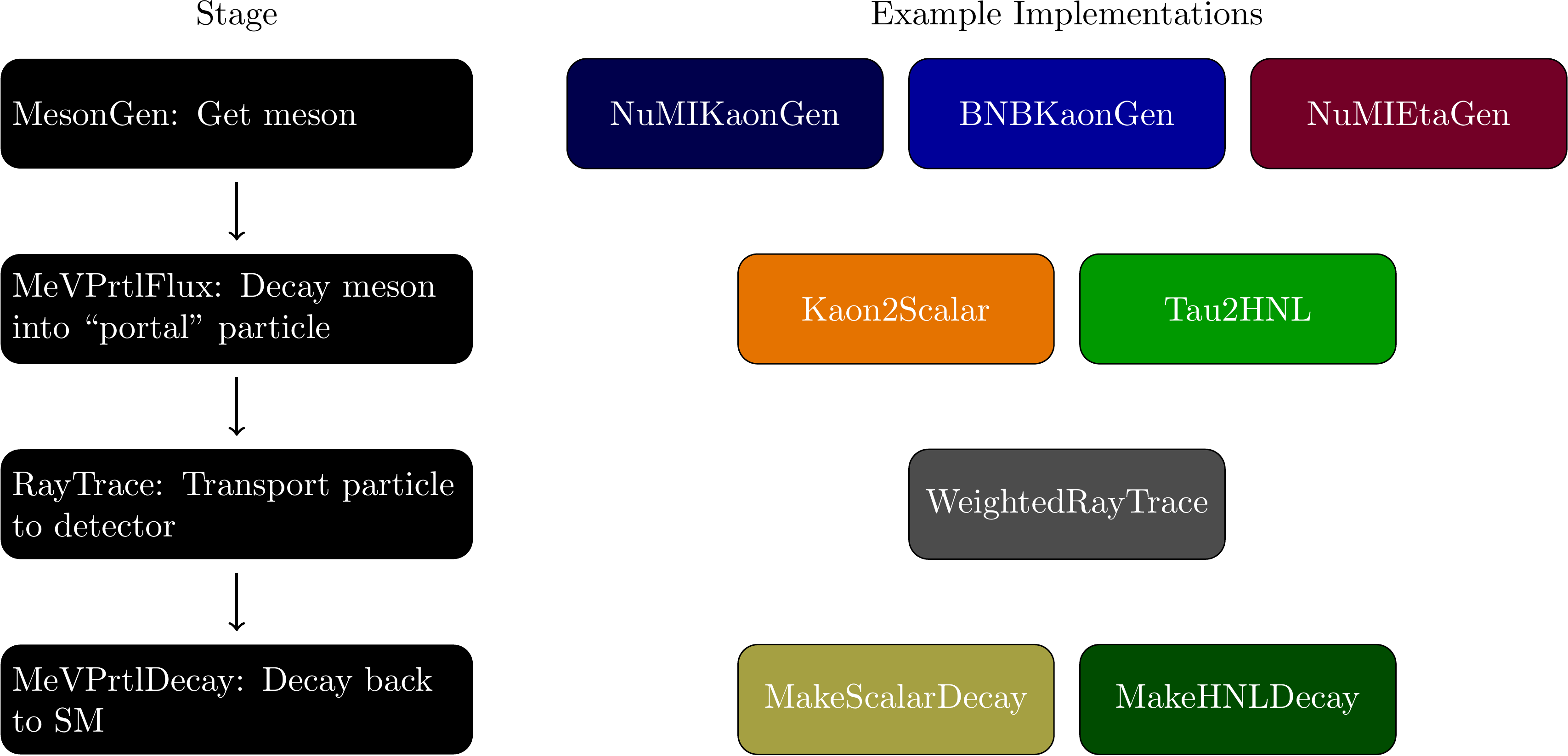}
    \caption{Schematic representation of the \lstinline{MeVPrtl} framework. Each of the four main stages of the generator are shown in consecutive order at left in black. Specific example implementations that could be run for each stage are shown in colored blocks to the right of the corresponding stage. There is exactly one implementation for the RayTrace stage, WeightedRayTrace, as that physics is common to any dark sector particle originating in the BNB or NuMI beam and traveling to the SBN Program detectors.}   
    \label{fig:tools}
    
\end{figure}

The \lstinline{MeVPrtl} generator simulates the production of BSM particles via meson decay or mixing in a neutrino beam, the BSM particles' decay in ICARUS or SBND~\cite{ICARUS:2004wqc,MicroBooNE:2015bmn,ICARUS:2023gpo}, and the associated probability or weight of each simulated event happening. As depicted in \cref{fig:mcgen}, \lstinline{MeVPrtl} is fully integrated with the SBN Program's \lstinline{LArSoft}-based \cite{Snider:2017wjd} simulation and reconstruction pipeline. Situated as the most upstream component of the pipeline, it can output an art-ROOT~\cite{Brun:1997pa} file containing the \lstinline{art} \cite{Green:2012gv} data products that are used as input for the \lstinline{Geant4}~\cite{Asai:2006qm,Allison:2006ve,Asai:2015xno} and detector response simulations. Alternatively, the generator can also output a ROOT \lstinline{TTree} for simple analysis of truth-level (non-reconstructed) events.

\lstinline{MeVPrtlGen} is the top-level module which implements the generator. From there, the generation process is split into four stages: extraction of mesons from provided flux files, determination of the portal particle flux, tracing portal particle trajectories from production to the detector, and determination of the decay in the detector. Each of the four stages is defined by a C++ interface: \lstinline{IMesonGen}, \lstinline{IMeVPrtlFlux}, \lstinline{IRayTrace}, and \lstinline{IMeVPrtlDecay}, respectively. By splitting the process into separate stages, different models can share some stages. For example, ray tracing is the same for all models. A diagram of the stages and some of their implementations is shown in \cref{fig:tools}. The physics for which each stage is responsible is discussed in detail in \cref{sec:simulation}.

The \lstinline{MeVPrtl} is hosted by the SBN Program on its GitHub repository for common analysis software, \lstinline{SBNSoftware/sbncode}. The generator's code can be found in \lstinline{sbncode/EventGenerator/MeVPrtl}. It is accessible on GitHub \cite{mevprtl_github_url}.
This description of \lstinline{MeVPrtl} describes the state of the generator as presented in \lstinline{sbncode} version \lstinline{v10_14_02_01}.

\section{Simulation Stages}
\label{sec:simulation}
This section provides an overview of the physics and implementation of the four stages involved in portal particle event generation within \lstinline{MeVPrtl}: \lstinline{IMesonGen}, \lstinline{IMeVPrtlFlux}, \lstinline{IRayTrace}, and \lstinline{IMeVPrtlDecay}.

\lstinline{MeVPrtl} can be configured to produce weighted or de-weighted events. Weighted event generation typically runs much more quickly (about one thousand times faster), but the output has a larger statistical uncertainty for the same number of generated events. It is therefore recommended to produce weighted events for quick, truth-level studies and de-weighted events for analyses requiring the full detector simulation. The event de-weighting is implemented by an accept-reject algorithm. The maximum possible weight $m$ is computed, and then a candidate event with weight $w$ is rejected with probability $1 - w/m$. This algorithm requires the maximum weight to be known a priori. Other event generators have techniques to estimate the maximum weight empirically. \lstinline{MeVPrtl} requires each stage to self-report its maximum possible weight based on the input configuration. This choice limits the event generator to simple production mechanisms for the visible final states (such as dark sector particle decays), which enable analytic determination of maximum weights.

The total weight corresponding to each event produced by \lstinline{MeVPrtl} is a product of contributions from various stages of generation, normalized by the number of protons on target (POT) generated versus expected: 

\begin{equation}
    w_{\rm total} = w_{\rm flux} \times w_{\rm ray} \times w_{\rm decay} \times \frac{\rm POT_{expected}}{\rm POT_{generated}}.
\label{Schematic Weight}
\end{equation}
$w_{\rm flux}$, $w_{\rm ray}$, and $w_{\rm decay}$ are discussed below in the context of the module in which they are calculated. $\text{POT}_{\text{generated}}$ is the number of simulated protons on target and $\text{POT}_{\text{expected}}$ is the number of protons on target expected for the experiment being studied. 

\Cref{subsec:MesonGen,subsec:MeVPrtlFlux,subsec:RayTrace,subsec:Decay} describe each of the above stages of event generation in detail.  Validation of the simulation is presented in \cref{sec:wgtVal}. \Cref{sec:timing} describes how event timing is handled in the generator.

\subsection{\lstinline{IMesonGen}: Meson Extraction}
\label{subsec:MesonGen}

The \lstinline{IMesonGen} class sources the parent particles, usually mesons, that will decay to or mix with the BSM particles, and returns their information in the same format as neutrino flux inputs currently utilized in the SBN Program's \lstinline{LArSoft}~\cite{Snider:2017wjd} analysis pipeline. The \lstinline{IMesonGen} class also returns the number of protons on target simulated for all of the provided parent particles since the last call. \lstinline{MeVPrtl} reuses the \lstinline{LArSoft} \lstinline{simb::MCFlux}~\cite{Church:2013hea,Pordes:2016ycs,Snider:2017wjd} object to encode the meson information to make it easier to incorporate existing neutrino simulation code. This object contains enough information on the mesons for the needs of \lstinline{MeVPrtl} model implementations. Typically the fields containing neutrino information are ignored. 

Information is extracted differently for long- and short-lived parent particles. To simulate long-lived mesons $K$ and $\pi^\pm$, the generator relies on the neutrino flux simulation in NuMI \cite{Adamson:2015dkw} and BNB \cite{Stancu:2001cpa}. Kaons are extracted from \lstinline{g4numi}~\cite{Wood:2024jos} and \lstinline{g4bnb}~\cite{MiniBooNE:2008hfu,MicroBooNE:2018efi,Paton:2025lmv} \lstinline{Geant4}~\cite{Asai:2006qm,Allison:2006ve,Asai:2015xno} simulation outputs in \lstinline{Dk2Nu} \cite{Hatcher:2012} and \lstinline{BooNE} format respectively.

Users are encouraged to check for kinematic cuts present in the \lstinline{Geant4} flux files, and remove or address them according to their physics model of interest.

The short-lived $\pi^0$, $\eta$, and $\eta'$ parent particles do not produce significant numbers of neutrinos and so are not kept in standard neutrino flux files.  On the other hand, these particles do not propagate significantly after production.  It is therefore a reasonable approximation to neglect the full structure of the beamline.  The conservative approximation of neglecting production in secondary interactions is employed.  The \lstinline{MeVPrtl} generator therefore uses \lstinline{Pythia8}~\cite{Bierlich:2022pfr} to simulate proton-nucleon interactions in the NuMI target. These collisions are simulated at the center-of-mass energy $E_{\text{cm}} = \sqrt{m_p^2 + m_N^2 + 2 m_N E_p}$, where $m_{N}$ is the mass of the target nucleon, $m_p$ is the mass of the incoming proton, and $E_p$ is the energy of the incoming proton. This comes out to $15.123~\text{GeV}$ for proton-proton collisions and $15.133~\text{GeV}$ for proton-neutron collisions at the target. 
The $E_{\text{cm}}$ approximation neglects nuclear motion, nuclear final state interactions, and secondary interactions. 
The collisions are generated with the \lstinline{SoftQCD} process, which simulates the low energy QCD interactions that dominate light meson production. The $\pi^0$, $\eta$, and $\eta^\prime$ momenta are recorded and used for generating dark sector fluxes. The meson counts per POT have been validated against results from Ref.~\cite{AxionSBN}. 

\subsection{\lstinline{IMeVPrtlFlux}: Portal Particle Flux Determination}
\label{subsec:MeVPrtlFlux}

The \lstinline{IMeVPrtlFlux} class handles the portal particle production via meson decay or mixing and sets the multiplicative weight associated with the production, $w_{\rm flux}$.

For portal particles originating from long-lived mesons ($M$), the flux weight takes the form 

\begin{equation}
    w_{\rm flux} = w_{\rm imp}\times\frac{\operatorname{Br}(M\to S + X)}{\operatorname{Br}(M\to \nu + X)}.
\end{equation}
The \lstinline{g4bnb} and \lstinline{g4numi} force meson decays into neutrinos, which is accounted for dividing by the SM branching fraction of meson to neutrinos.  $\operatorname{Br}(M\to S + X)$ is then included to account for the probability that the meson will decay to the portal particle of interest.  The importance weight~\cite{MiniBooNE:2008hfu} $w_{\rm imp}$ is a statistical weight output by \lstinline{g4bnb} or \lstinline{g4numi}, reflecting the modification of used cross sections (\lstinline{g4bnb}) or encoding the survival probability of a parent particle (\lstinline{g4numi}). 
For portal particles that come from short-lived mesons, $w_{\rm flux} = \operatorname{Br}(M\to S + X)$. If it is impossible for a given meson to produce a portal particle, the meson is skipped.

The \lstinline{IMeVPrtlFlux} class also computes and returns the largest possible weight the class would assign to an event. This allows the module to de-weight events according to the accept-reject algorithm.

\subsection{\lstinline{IRayTrace}: Ray Tracing}
\label{subsec:RayTrace}

The \lstinline{IRayTrace} class takes as input the portal particle flux, and outputs a direction such that the particle impinges a specified detector volume, along with the associated weight $w_{\rm ray}$.

The geometry of the detector is modeled as a box containing the active volume(s) of the detector.  The dimensions are specified in a coordinate system where $z$ is the BNB axis and $y$ is vertical.  In the case of a NuMI-generated flux, the detector is then rotated to a coordinate system where $z$ is the NuMI beam axis.  The coordinates of the center of the detector are specified in the relevant beam frame for the given simulation, with $\theta$ being the polar angle with respect to the beam axis and $\phi$ being the azimuthal angle around the beam axis.  This frame is referred to as the lab frame below.

To ensure the event generation runs efficiently, it is necessary to force the portal particle ray to hit a detector volume. This is done by picking that particle's direction such that it impinges the volume, and weighting the event by the probability that the particle traveled in the direction of the detector, $w_{\rm ray}$. In the case of a particle produced isotropically in the decay of a parent particle, this probability is equal to 
\begin{equation}
    w_{\rm ray} = \frac{1}{4\pi}\oiint\limits
    d \Omega^\prime \, ,
\end{equation}
where $\Omega^\prime = (\theta^\prime, \phi^\prime)$ is the solid angle in the rest frame of the meson parent. The detector surface area is not easily expressible in this frame, so the integral is done using a change of variables,
\begin{equation}
\label{eq:probabilityintegral}
    w_{\rm ray} = \frac{1}{4\pi}\oiint\limits 
    \left|\frac{d\Omega^\prime}{d\Omega}\right| d \Omega \, ,
\end{equation}
where $\Omega = (\theta, \phi)$ is the lab frame solid angle consisting of polar angle $\theta$ with respect to the beam direction and azimuthal angle $\phi$. The event generator implements this integral by picking the lab frame direction $\Omega$ and applying the Jacobian $|d\Omega^\prime/d\Omega|$ as a weight on the event. This weight is known for the neutrino case, where the child particle is approximately massless \cite{MINOS}. Here, the result for a massive particle is derived.

The known inputs are the momentum and energy of the child particle, assuming a two-body decay process to produce it, in the parent rest frame ($p^\prime$ and $E^\prime$) and the angle of the child particle with respect to the parent particle direction in the lab frame $(\theta, \phi)$ -- picked so that the child particle hits a specified volume. Then, the remaining kinematics $p$ (the child particle momentum in the lab frame) and $\theta^\prime$ (the polar angle of the child particle in the parent particle's frame) are defined by
\begin{equation}
    \begin{split}
    \label{eq:labframemomentum}
        \cos\theta'_\pm &= \frac{-E'\beta\gamma\text{sin}^2\theta \pm \cos\theta\sqrt{E'^2/\gamma - m^2 + m^2\beta^2 \cos^2\theta}}{p'\gamma (1-\beta^2\cos^2\theta)},\\
        p_\pm &= \frac{E'\beta\cos\theta/\gamma \pm \sqrt{E'^2/\gamma^2 - m^2 + m^2 \beta^2 \cos^2\theta}}{1 - \beta^2\cos^2\theta} \, ,
    \end{split}
\end{equation}
where $\beta$ and $\gamma$ are the usual quantities from special relativity given by the velocity and boost of the parent particle in the lab frame and $m$ is the child particle mass. Unlike the massless (neutrino) case where there is always exactly one solution to this problem ($\frac{d\Omega'}{d\Omega} = E^2/E'^2$), in the massive case there can be zero, one, or two solutions. The ($\pm$) solution exists when $p_\pm$ is positive. The weight is then equal to 
\begin{equation}
    \begin{split}
    \label{eq:rayforceweight}
        \left(\frac{d\Omega^\prime}{d\Omega}\right)_\pm =& \frac{E^\prime}{E_\pm}\frac{p_\pm^2}{p^{\prime 2}}\frac{dp}{dp^\prime}_\pm, \\
        \left(\frac{dp}{dp^\prime}\right)_\pm =& \frac{p^\prime}{E^\prime\gamma} \frac{\beta\cos\theta \pm E^\prime / \sqrt{E^{\prime 2} - \gamma^2 m^2 (1-\beta^2\cos^2\theta)}}{1 - \beta^2 \cos^2\theta}\, .
    \end{split}
\end{equation}
This weighting scheme is used in the \lstinline{WeightedRayTraceBox} implementation of \lstinline{IRayTrace}, which forces $\theta$ and $\phi$ to hit the detector volume and computes the (unbounded) weight.

\begin{figure}[!tbh]
    \centering
    \includegraphics[width=0.45\textwidth]{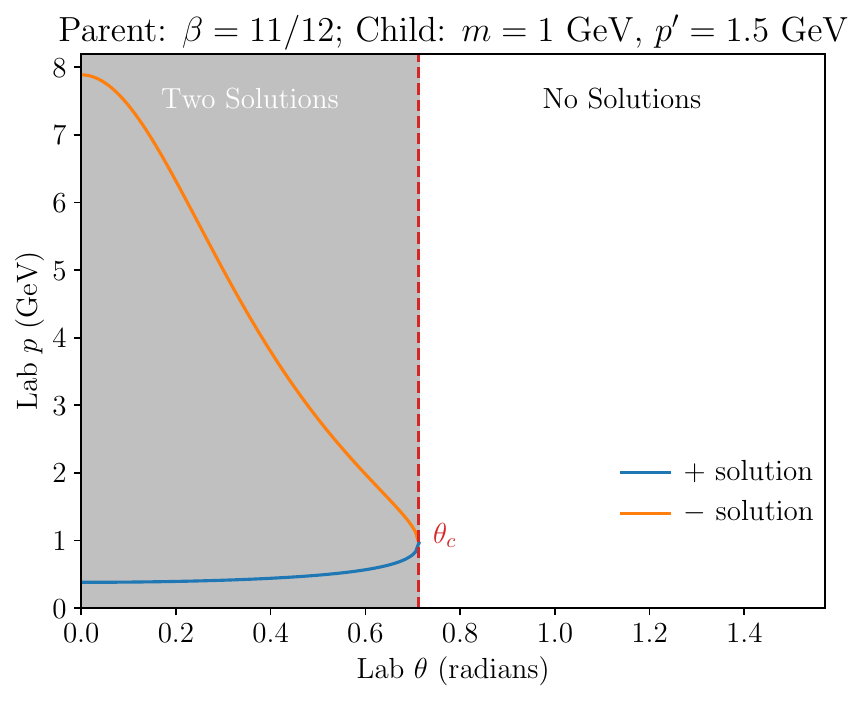}
    \includegraphics[width=0.45\textwidth]{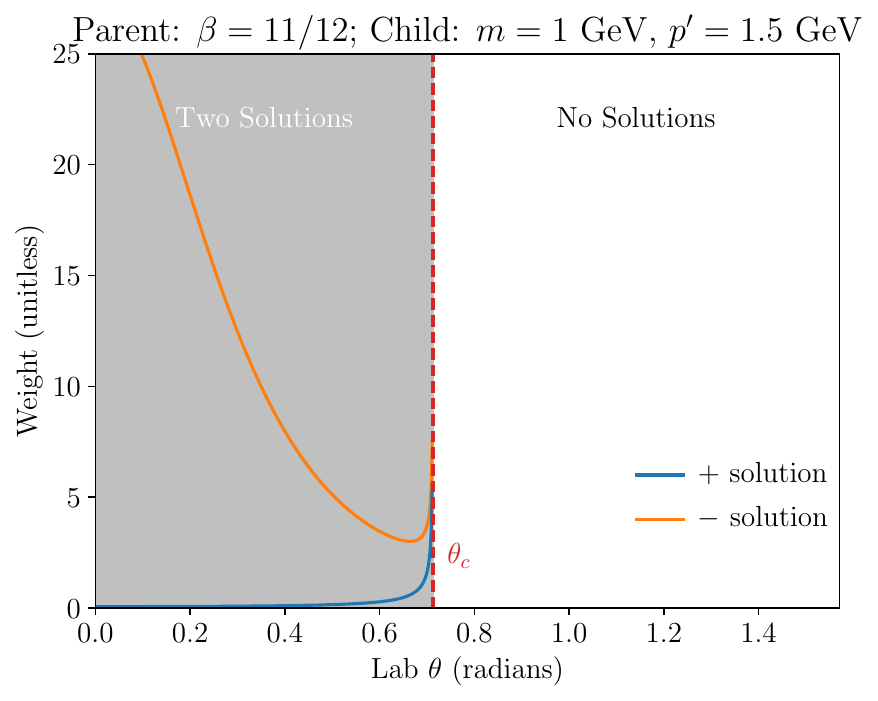}
    \caption{Example solutions to the lab frame momentum (\cref{eq:labframemomentum}) and the ray-tracing weight (\cref{eq:rayforceweight}). The solutions are specified by the parent velocity in the lab frame ($\beta$) and the child mass ($m$) and momentum in the parent rest frame ($p^\prime$).}   
    \label{fig:rayforcesol}
\end{figure}
Example numerical solutions for the weight and the lab frame momentum are shown in \cref{fig:rayforcesol}. When the speed of the parent particle in the lab frame is greater than the speed of the massive child particle in the rest frame, there is a critical angle, $\cos{\theta_c} = \sqrt{m^2 - E^{\prime 2}/\gamma^2}/(m\beta)$, which is the largest possible angle the child particle can attain. The corresponding (integrable) ray weight at the critical angle is infinite, meaning that de-weighting with an accept-reject algorithm based on the maximum possible weight cannot be performed. 

To overcome this issue, \lstinline{MeVPrtl} provides an alternative computation that only forces the $\phi$ angular variable to hit the detector, while still picking $\theta^\prime$ randomly. In this case, the relevant Jacobian is just $d\phi^\prime/d\phi = 1$, and the weight is scaled by the fraction of successful throws of $\theta^\prime$. The event generation is still reasonably efficient ($\gtrsim 0.1$\% of throws are successful) with only one angular variable forced to intersect the detector volume.  This scheme is used in the \lstinline{MixedWeightRayTraceBox} implementation of \lstinline{IRayTrace}, which only forces $\phi$ to impinge the detector volume and computes the maximum weight.  This module also computes a (configurable) number of random throws of $\theta^\prime$ to increase the odds that at least one corresponding $\theta$ will impinge the detector. 

If forcing the ray to hit the detector is not possible, \lstinline{IRayTrace} drops the event. The \lstinline{IRayTrace} class also returns the largest possible $w_{\rm ray}$ the class would assign to an event. 

\begin{figure}[!tbh]
    \centering
    \begin{subfigure}{0.45\textwidth}
        \centering
        \textbf{WeightedRayTraceBox Validation}
        \includegraphics[width=\textwidth]{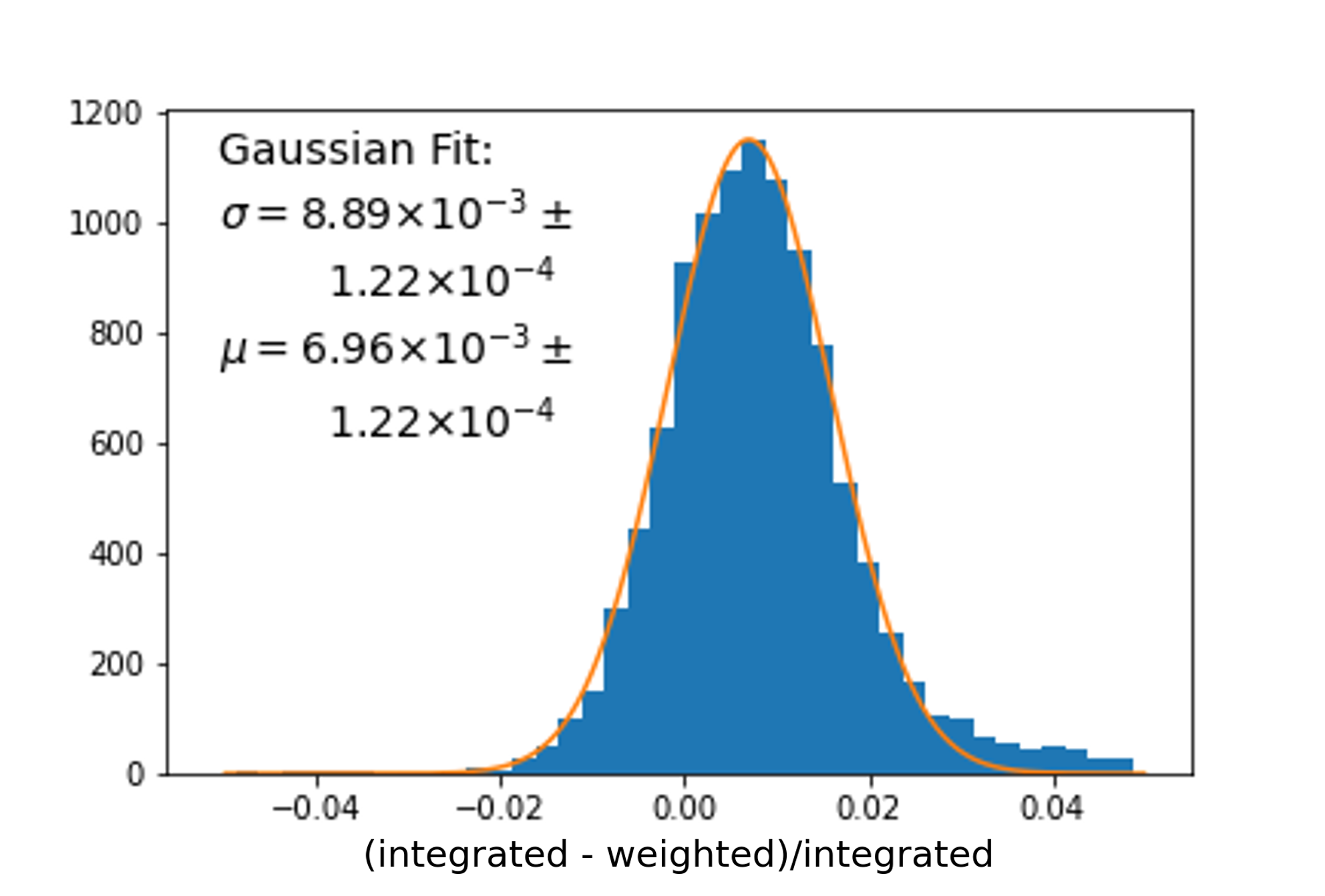}
    \end{subfigure}
    \begin{subfigure}{0.45\textwidth}
        \centering
        \textbf{MixedWeightRayTraceBox Validation}
        \includegraphics[width=\textwidth]{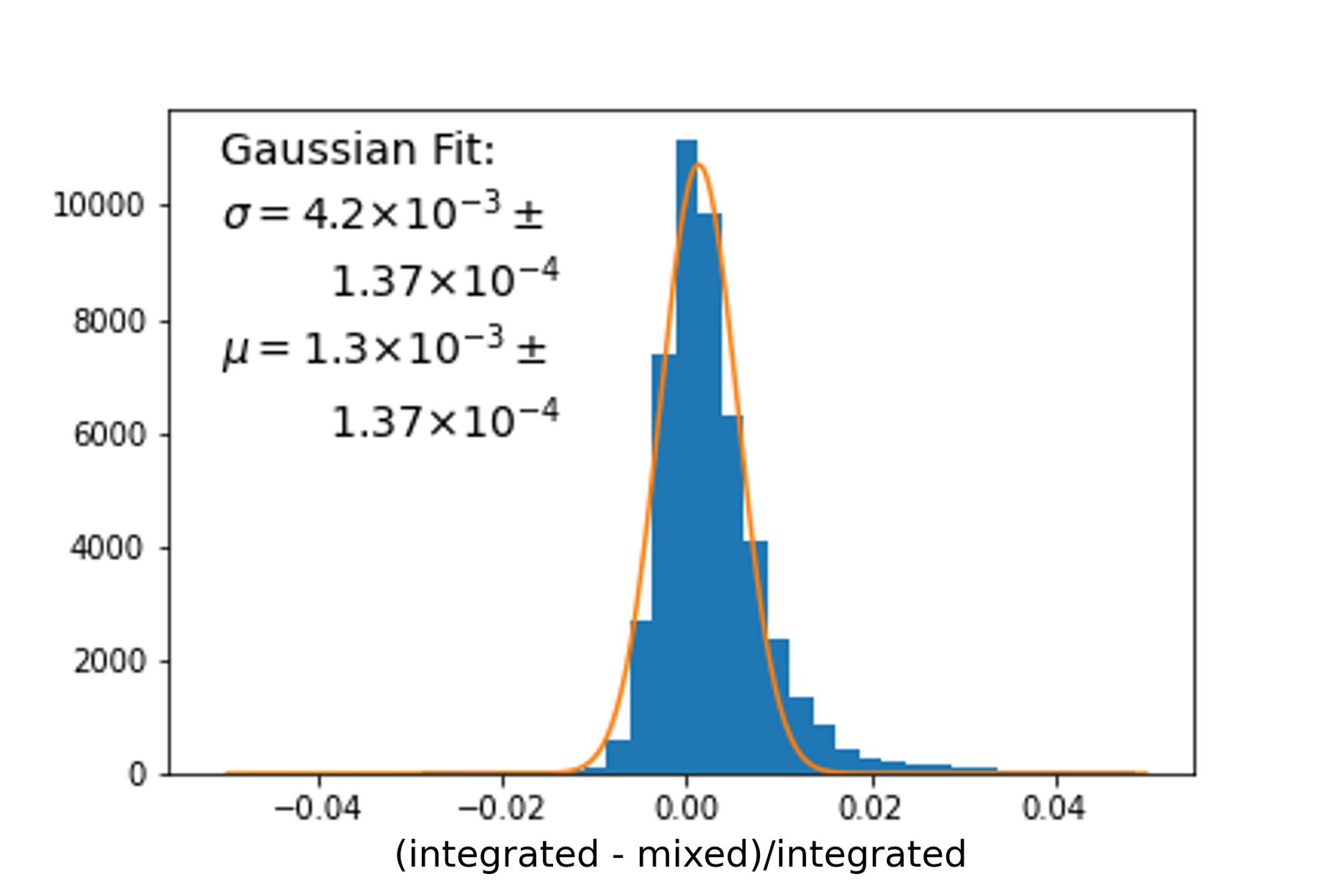}
    \end{subfigure}
    \caption{Comparison of the probability of intersection computation from numerically integrating \cref{eq:probabilityintegral} (labeled as integrated) and the two Monte-Carlo-based methods: (left) \lstinline{WeightedRayTraceBox} (labeled as weighted) and (right) \lstinline{MixedWeightRayTraceBox} (labeled as mixed). The comparison is done for the HPS model.}  
    \label{fig:rayforcevalidation}
\end{figure}

The ray tracing methods -- both \lstinline{MixedWeightRayTraceBox} and \lstinline{WeightedRayTraceBox} -- were validated by comparing the weights from the two Monte Carlo methods to a numerical computation of the probability weight integral (\cref{eq:probabilityintegral}). As seen in \cref{fig:rayforcevalidation}, both methods reproduce the central value of the integral to better than a percent. There is a spread in the comparison due to the stochastic nature of the two Monte Carlo methods. This spread is smaller for the mixed-weight method.  For this reason, the default, recommended ray tracing method is \lstinline{MixedWeightRayTraceBox}.

\subsection{\lstinline{IMeVPrtlDecay}: Portal Particle Decay}
\label{subsec:Decay}

The \lstinline{IMeVPrtlDecay} class also takes the portal particle flux as input, and outputs the decay information of the particle. This information includes the portal particle's decay position and time, and the list of particles it decays to. The associated probability for the particle decaying to the final state of interest in a detector, $w_\text{decay}$, is given by 
\begin{equation}
   w_\text{decay} = (e^{-L_\text{enter}/\gamma \beta \tau} - e^{-L_\text{exit}/\gamma \beta \tau}) \times \operatorname{Br}(S \to \text{final state})
\end{equation}
where $\gamma$ and $\beta$ are the Lorentz boost and velocity for the portal particle ($S$) with respect to the lab frame and $\tau$ is its proper lifetime. $L_\text{enter}$ and $L_\text{exit}$ correspond to the points along the particle's trajectory where, if the particle is stable enough, it would enter and exit the detector. ``Final state'' refers to the decay products being simulated (e.g.\ $\mu^+\mu^-$ for an HPS decaying into the dimuon channel). 
Like the other stages of \lstinline{MeVPrtl}, the class also returns the largest possible $w_\text{decay}$ the class would assign to an event. 

\subsection{Event Weight Validation}
\label{sec:wgtVal}

\lstinline{MeVPrtl}'s determination of event weights has been validated in the context of HPS generation at ICARUS by an independent approach described by the following formalism.  For an overview of this model, see \cref{sec:models_higgsportal} below, as well as Ref.~\cite{Pospelov:2007mp}.

For a given event, the differential probability of HPS decay in terms of its decay position relative to the parent kaon is given by 
\begin{equation}
    dP = \frac{1}{4\pi} 
    \frac{1}{\gamma v \tau} e^{-r/\gamma v \tau}
    dr d\cos\theta' d\phi',
\label{Higgs_dPprime}
\end{equation}
where $r$ is now the length of the HPS trajectory, $v$ is the HPS speed, and $\tau$ is the HPS lifetime. In terms of ICARUS detector coordinates $x_{d}, y_{d}, z_{d}$ this becomes 

\begin{equation}
    dP = \frac{1}{4\pi} 
    \frac{1}{\gamma v \tau} e^{-r(x_d,y_d,z_d)/\gamma v \tau}
    \left \lvert \frac{\partial(\cos\theta^\prime,\phi^\prime)}{\partial(\cos\theta,\phi)} \right \rvert
    \left \lvert \frac{\partial(r,\cos\theta,\phi)}{\partial(x_{d},y_{d},z_{d})} \right \rvert
    dx_{d} dy_{d} dz_{d}. 
\label{Higgs_dP}
\end{equation}
The first Jacobian is the same as $\left \lvert d\Omega^\prime/d\Omega \right \rvert$ in \cref{eq:rayforceweight}.
The second takes the form $1/r^2$, following the usual transformation from a spherical to Cartesian system. For any given event, the scalar's trajectory $r = r(x_d,y_d,z_d)$ is well defined by the scalar and kaon decay positions. 

Event weights are defined such that a sum over the event weights in a sample of weighted events, to be normalized by the POT as explained in \cref{Schematic Weight}, gives the number of expected interactions in the detector ($N'$). This statement assumes the POT is exactly known. Monte Carlo generation performs this sum via implicit integration of the likelihood of each event, over the detector volume. Leaving the POT-normalization to be accounted for downstream, the kaon importance weight, relevant branching fractions, and differential probability of decay from \cref{Higgs_dP} contribute to the likelihood of each event:

\begin{equation}
    N' = 
    \int_{\substack{\text{detector} \\ \text{volume}}}{
        w_{\text{imp}}\frac{\Gamma(K^\pm\rightarrow\pi^\pm S)}{\Gamma(K^\pm\rightarrow\nu X)}
        \frac{\Gamma(\textrm{S} \rightarrow \textrm{final state})}{\Gamma(\textrm{S} \rightarrow \textrm{any})}
        \frac{dP}{dx_{d} dy_{d} dz_{d}}
        dx_{d} dy_{d} dz_{d}.
    }
\label{Nexpected_int}
\end{equation}

In accordance with the mean value theorem, the integral can be approximated as a product of the detector volume and a sum over the form of the integrand for each generated event, normalized by the number of generated events in the sample ($N$):

\begin{equation}
    N' = 
    \frac{V_{\text{detector}}}{N}
    \sum_{i=1}^{N}{
        {w_{\text{imp}}}_{i}\frac{\Gamma(K^\pm\rightarrow\pi^\pm S)}{\Gamma(K^\pm\rightarrow\nu X)}
        \frac{\Gamma(\textrm{S} \rightarrow \textrm{final state})}{\Gamma(\textrm{S} \rightarrow \textrm{any})}
        \left(\frac{dP}{dx_{d} dy_{d} dz_{d}}\right)_i.
    }
\label{Nexpected_sum}
\end{equation}

Further leaving $N$ to be re-accounted for downstream, and distributing the factor of the detector volume into the form of the integrand for each event, the total weight of a given event is given by

\begin{equation}
    w_{\text{total}} = w_{\text{imp}}\frac{\Gamma(K^\pm\rightarrow\pi^\pm S)}{\Gamma(K^\pm\rightarrow\nu X)}
    \frac{\Gamma(\textrm{S} \rightarrow \textrm{final state})}{\Gamma(\textrm{S} \rightarrow \textrm{any})}
    V_{\text{detector}}
    \frac{dP}{dx_{d} dy_{d} dz_{d}}.
\label{HiggsValidationWeight}
\end{equation}

This formalism lends itself to sampling HPS decay positions from the ICARUS detector volume. Indeed, this is done in practice for the validation approach. As long as the scalar's velocity in the kaon rest frame, $v^\prime_S$, is greater than the kaon's lab-frame velocity $v_K$, sampling scalar decay positions in the detector volume will result in a finite maximum weight. With the scalar decay position fixed entirely by sampling, though, it remains possible that the singularity in $\left \lvert \frac{d\Omega^\prime}{d\Omega} \right \rvert$ is encountered when $ v_K > v^\prime_S $ (as discussed in \cref{subsec:RayTrace}).
In the validation approach, event de-weighting is thus performed with an accept/reject algorithm that uses the maximum \textit{generated} weight in the sample, in contrast to the maximum possible weight as is done by \lstinline{MeVPrtl}. While one must be extraordinarily unlucky to sample and generate an event with kinematics landing exactly on the singularity, at high statistics, generating at least one event near
the singularity (i.e. on the asymptote to the infinite weight) is not impossible. Since encountering a maximum weight near the singularity any time the generator is run is left to chance, leading to unpredictable and possibly very poor de-weighting efficiency, this method is reserved for validation purposes only.

\begin{figure}[!tbh]
    \centering
    \begin{subfigure}{0.47\textwidth}
        \centering
        \includegraphics[width=\textwidth]{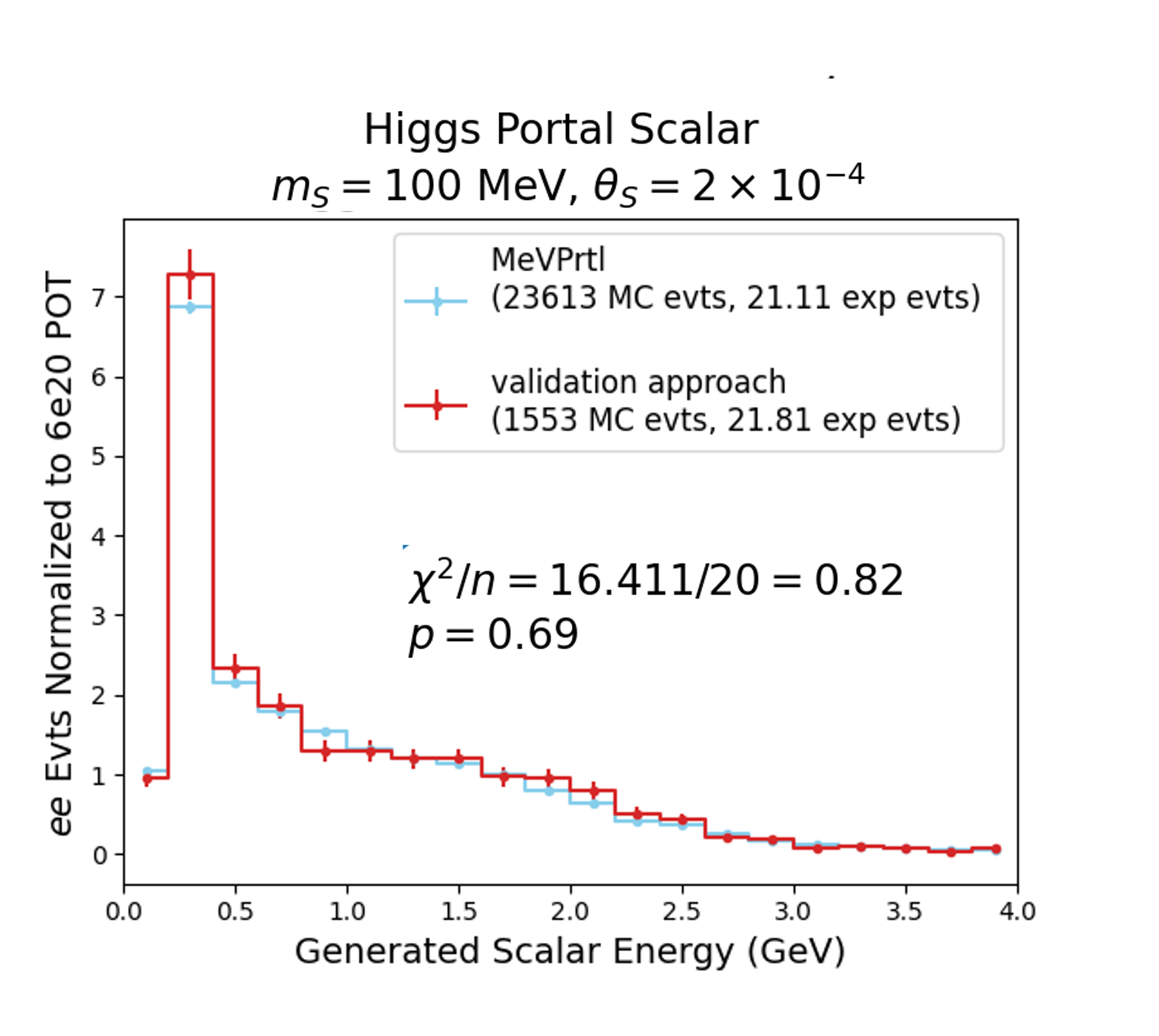}
    \end{subfigure}
    \begin{subfigure}{0.45\textwidth}
        \centering
        \includegraphics[width=\textwidth]{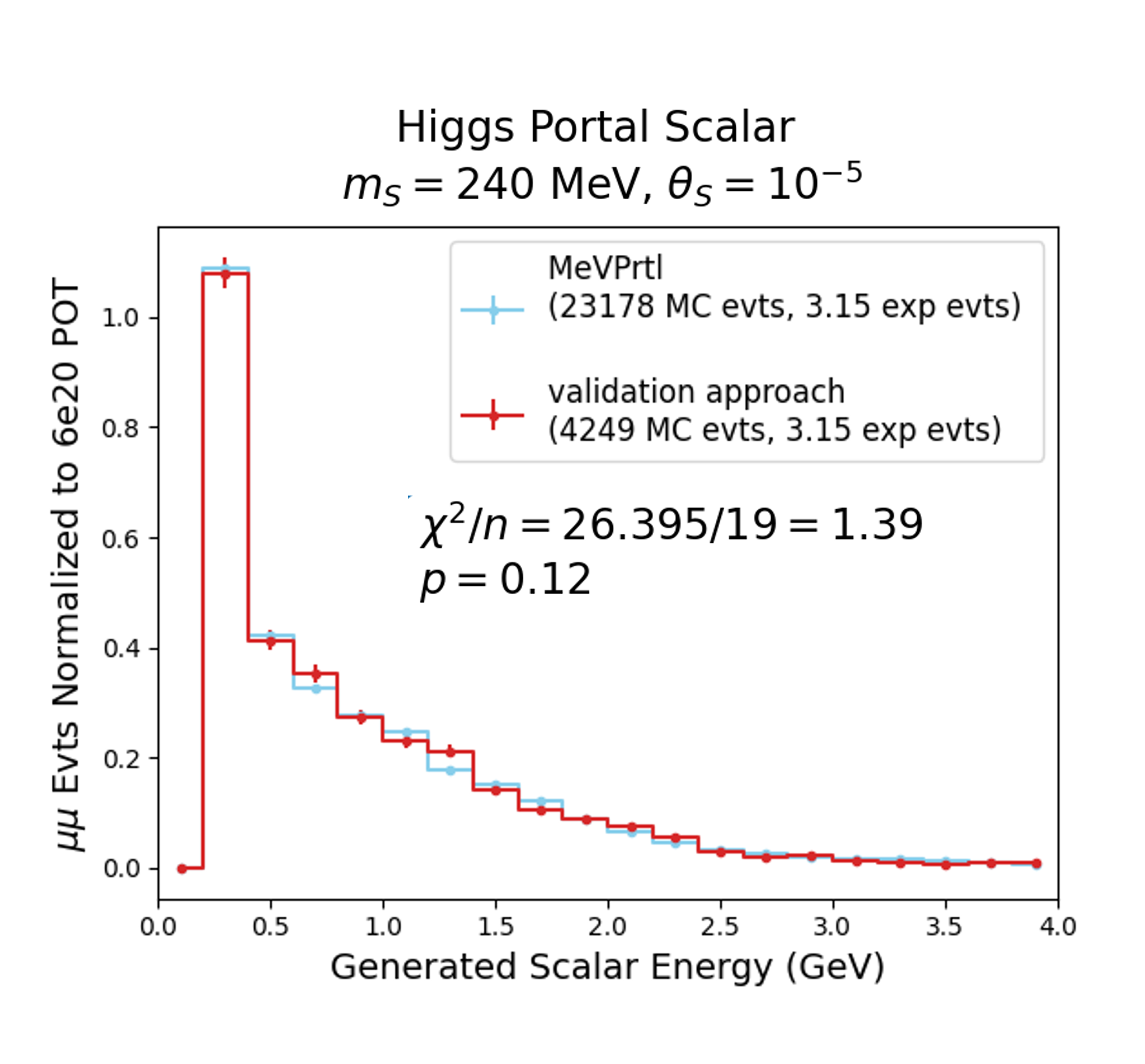}
    \end{subfigure}
    \begin{subfigure}{0.45\textwidth}
        \centering
        \includegraphics[width=\textwidth]{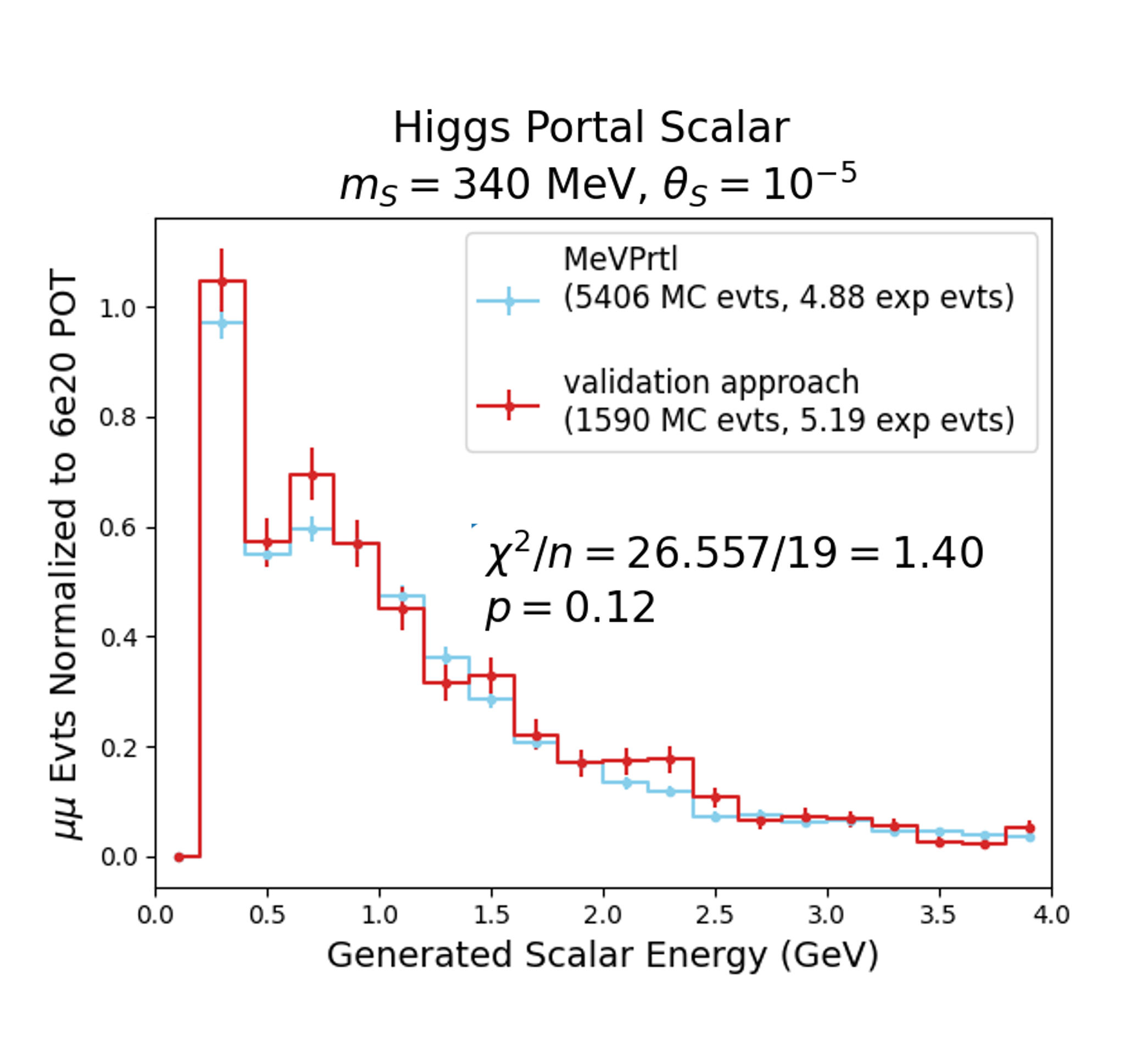}
    \end{subfigure}
    
    \caption{Demonstrated agreement between the \lstinline{MeVPrtl} generator and the validation approach for three different HPS model benchmarks. The spectra generated by \lstinline{MeVPrtl} show consistency with those generated with the validation approach for all three model benchmarks, with p-values for $\chi^2$ given $n$ equal to 0.69, 0.12, 0.12 for the $m_S = 100~ \text{MeV}$ with $\theta_S = 2 \times 10^{-4}$, $m_S = 240~\text{MeV}$ with $\theta_S = 10^{-5}$, and $m_S = 340~\text{MeV}$ with $\theta_S = 10^{-5}$ HPS benchmarks respectively.}     
    \label{fig:HPSval}
\end{figure}

To validate that \lstinline{MeVPrtl} is correctly weighting HPS events, the expected event rates and energy spectra of HPS generated by \lstinline{MeVPrtl} to samples generated with the validation approach have been compared. The validation was performed for three different model benchmarks: $m_S = 100~ \text{MeV}$ with $\theta_S = 2 \times 10^{-4}$, $m_S = 240~\text{MeV}$ with $\theta_S = 10^{-5}$, and $m_S = 340~\text{MeV}$ with $\theta_S = 10^{-5}$. The parameter $\theta_S$ refers to the HPS mixing angle, which is described in detail in \cref{sec:models_higgsportal}. For the 240 MeV and 340 MeV mass benchmarks, samples consisting only of scalars that decay to muons are compared. This comprises the vast majority of 240 MeV scalars, since pion final states are not kinematically allowed there and the lepton decay width scales with the lepton mass squared. Decay to pions dominates for 340 MeV scalars, but a comparison for the fraction of scalars that decay to two muons verifies correct handling of relative branching fractions in the generators. \Cref{table:higgs_val_samples} shows the number of de-weighted Monte Carlo simulation events for each sample used for the comparison, along with the corresponding projected number of events to occur in ICARUS for $6\times10^{20}$ protons on target from the NuMI beam (roughly equal to ICARUS's exposure to NuMI during its first three runs of data collection).  The number of events expected according to each generation method agrees within the statistical uncertainty, thus validating the approach employed in \lstinline{MeVPrtl}.  The POT-normalized spectra are shown in \cref{fig:HPSval}. The statistics and event rates in the legends in \cref{fig:HPSval} reflect only the portions of each sample included in the plotted range (0--4 GeV). A $\chi^2$ comparison is computed for each pair of spectra over the plotted range, so that the number of degrees of freedom $n$ for each fit is equal to the number of non-empty bins. The spectra generated by \lstinline{MeVPrtl} show consistency with those generated with the validation approach for all three model benchmarks, with p-values for $\chi^2$ given $n$ shown in \cref{table:higgs_val_samples}.

\begin{table}[h!]
\begin{center}
    \begin{tabular}{|p{5cm}|p{3cm}|p{3cm}|p{1.25cm}| }   
        \hline
        \textbf{HPS model benchmark (final state)}  & \textbf{Events (\lstinline{MeVPrtl})} & \textbf{Events (Validation)} & p-value\\ 
        \hline 
        \hline
        {$m_S = 100~ \text{MeV}$, $\theta_S = 2\times10^{-4}$ ($ee$)} & 21.41 & 22.09 & 0.69\\
        \hline 
        {$m_S = 240~ \text{MeV}$, $\theta_S = 10^{-5}$ ($\mu\mu$)} & 3.19 & 3.19 & 0.12 \\
        \hline 
        {$m_S = 340~ \text{MeV}$, $\theta_S = 10^{-5}$ ($\mu\mu$)} & 5.05 & 5.32 & 0.12\\
        \hline
    \end{tabular}
    \caption{The number of generated and POT-normalized HPS events occuring in ICARUS according to \lstinline{MeVPrtl} and the validation approach for three different benchmark scalar masses. The events generated are normalized to $6\times 10^{20}$ protons on target. The p-value for $\chi^2$ given $n$ for each model benchmark is also shown here.}
    \label{table:higgs_val_samples}
\end{center}
\end{table}

\subsection{Event Timing}
\label{sec:timing}

\lstinline{MeVPrtl} calculates portal particle event timestamps so that they are comparable to the timestamp of a neutrino event produced by the \lstinline{GENIE} neutrino generator \cite{Andreopoulos:2015wxa}. The event timing begins when the first proton arrives at the target ($t=0$).  Three processes as described below contribute to the total event timing.

The beam spill time contribution to the event timing takes into account the unique BNB and NuMI beam structures. 
It is implemented in the production class \lstinline{IMeVPrtlFlux::BeamOrigin}, which randomly picks a point in the relevant beam spill window, and adds a global offset if required.

The beam spill time contribution is supplemented by the time that it takes for the parent meson to travel from its production position to its decay position, which includes its time of flight and other complex processes such as re-scattering. For the long-lived mesons $K^\pm$, $K^0_L$, and $\pi^{\pm}$, this time component is directly extracted from the flux file (\lstinline{dk2nu} and \lstinline{BooNE} flux for the NuMI and BNB beam respectively). The meson time of flight is negligible for the short-lived mesons simulated with \lstinline{Pythia8}. 

The final contribution comes from the time taken for the portal particle to travel from its production position to its decay position, given its initial energy and total travel distance. 

\section{Specific Dark Sector Model Implementations}
\label{sec:models}
There is a shared general phenomenology to the production and detection of BSM particles at neutrino experiments, where they are produced in the neutrino beam, propagate to a detector, and decay into SM particles. However, model specifics determine the portal particles' branching ratio at production and decay widths to SM final states. The physics and practical implementations specific to each of the models presently in \lstinline{MeVPrtl} are described below.

\subsection{Higgs Portal Scalar}
\label{sec:models_higgsportal}
In the HPS model, the mediator between the SM and dark sector is a neutral scalar particle ($S$) that mixes with the Higgs boson. This behavior is generic in models with new neutral scalars, and is one of the three renormalizable portals~\cite{Pospelov:2007mp}. The model is described in the mass basis by the interaction Lagrangian
\begin{equation}
    \mathcal{L}_{\rm int} = \sin\theta_S \, S\left(\frac{2 m_W^2}{v} \, W^{+\mu} W^-_\mu + \frac{m_Z^2}{v} \, Z^\mu Z_\mu - \sum_f \frac{m_f}{v}\,  \overline{f} f\right),
\end{equation} 
where $\theta_S$ is the mixing angle of the scalar $S$ with the Higgs, $v$ is the Higgs vacuum expectation value, and $f$ are the SM fermions. The parameter space is in this way characterized by the mass of the scalar $m_S$ and the mixing angle $\theta_S$. Decay products of mediators with masses below the K-$\pi$ mass difference and mixing angles down to order $10^{-4}$ can be detected by the SBN Program LArTPCs. In particular, scalars produced from NuMI kaons and decaying to $e^+e^-$, $\mu^+\mu^-$, or a pair of pions in ICARUS have the greatest projected sensitivity in the range $m_S \approx 40-360~\text{MeV}$ \cite{PhysRevD.100.115039}. Searches at MicroBooNE have demonstrated sensitivity to the HPS parameter space~\cite{uB_hps_and_hnl_2022}. 

Scalar production and decay within \lstinline{MeVPrtl} are described below, followed by discussions on special considerations for kaon kinematics at ICARUS (which is projected to have leading sensitivity to this model). An example for how to generate HPS events with \lstinline{MeVPrtl} is presented in \cref{apn:runHPS}.

\subsubsection{Production}

\begin{figure}
    \centering
    \begin{subfigure}{0.4\textwidth}
        \centering
        \textbf{}
        \includegraphics[width=\textwidth]{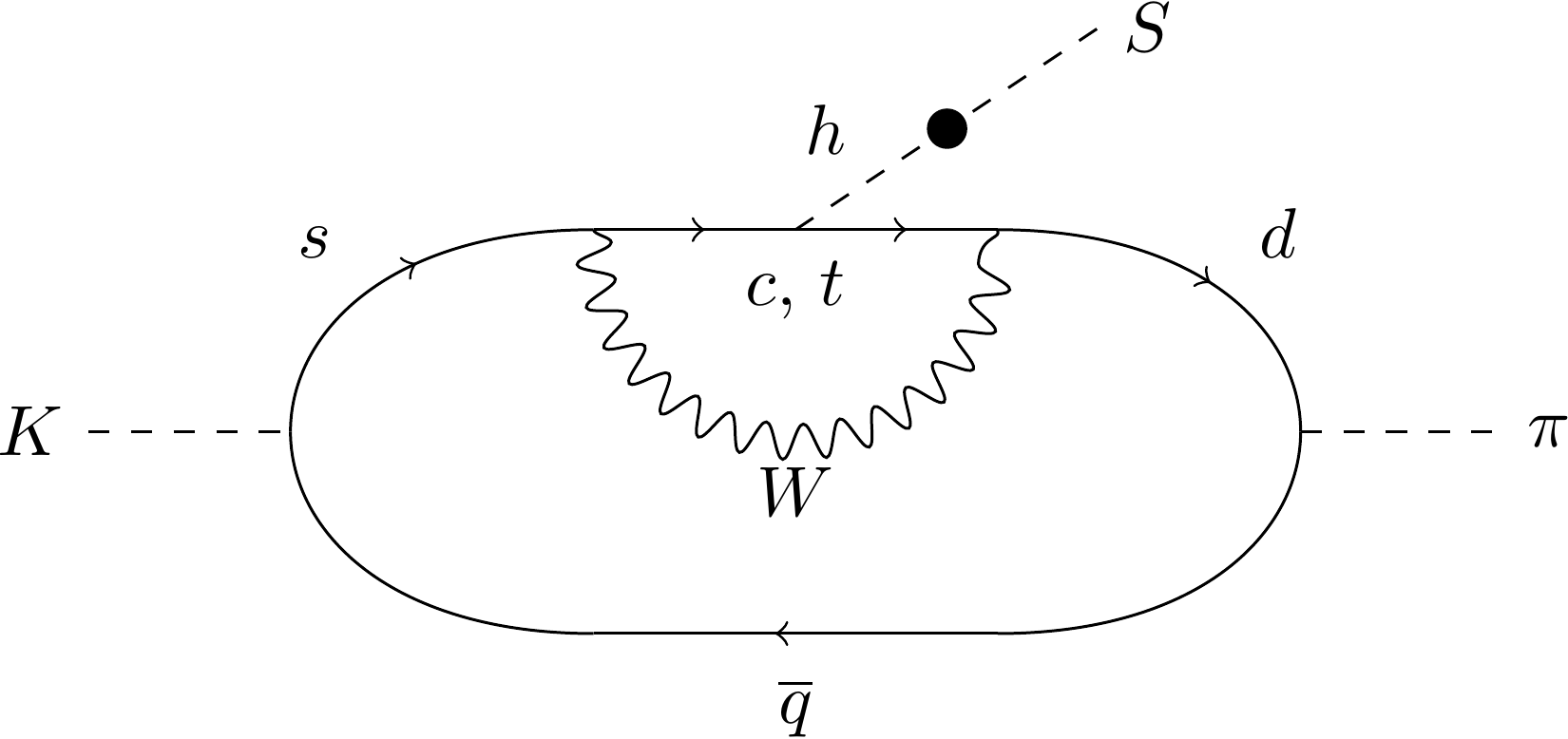}
    \end{subfigure}
    \hspace{1cm}
    \begin{subfigure}{0.27\textwidth}
        \centering
        \textbf{}
        \includegraphics[width=\textwidth]{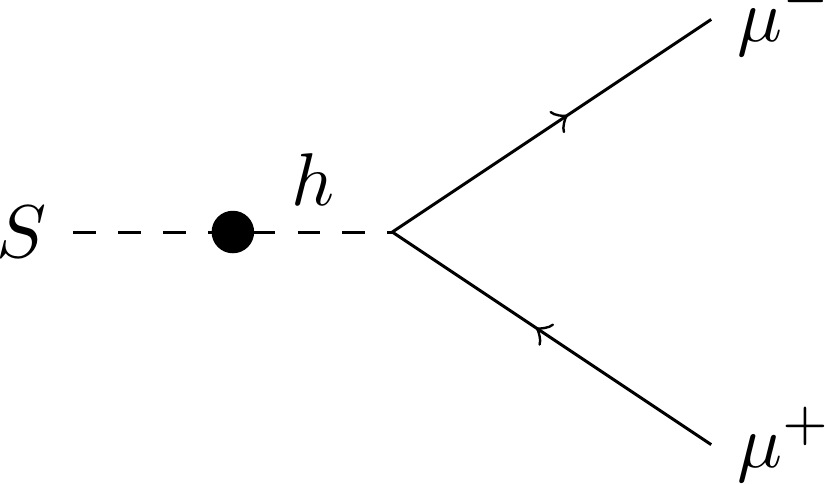}
    \end{subfigure}
    \caption{\textbf{Left:} HPS ($S$) production from kaons. \textbf{Right:} HPS decay to two leptons, in this case muons. The filled dots represent mixing between $S$ and the Higgs boson ($h$).}   
    \label{fig:hpsFeynmanDiagrams}
\end{figure}

Kaon decay $K\rightarrow \pi S $ is the dominant production channel for HPSs at the SBN Program \cite{PhysRevD.100.115039}, where $K$ can be either $K^\pm$ or $K^0_L$. The leading diagram, shown in the left panel of \cref{fig:hpsFeynmanDiagrams}, for this process is a W boson-top quark penguin diagram with partial width

\begin{equation}
    \operatorname{Br}(K\to \pi S) = \frac{\theta_S^2 \tau_K}{16\pi m_K} \left\lvert\frac{3V_{\rm CKM}m_t^2m_K^2}{32\pi^2 v^3}\right\rvert^2 \frac{2p_S}{m_K},
\label{HiggsProductionBr}
\end{equation} 
where $v$ is the Higgs vacuum expectation value and $V_{CKM}$ are elements from the CKM quark-mixing matrix \cite{PhysRevD.100.115039}. $V_{\rm CKM}$ = $V_{td} \times V_{ts}$ for $K^\pm$, and $Re(V_{td}) \times V_{ts}$ for $K^0_L$.
Since HPS production via $K^0_S$ is significantly smaller, only $K^\pm$ and $K^0_L$ parents are considered in \lstinline{MeVPrtl}. As described in \cref{subsec:MesonGen},
these kaons are extracted from neutrino flux files, then re-decayed into HPSs with branching ratios rescaled according to eq.~\eqref{HiggsProductionBr}.

Scalars may also be produced via B meson decays and proton bremsstrahlung processes \cite{Scalar_production_channels}. Very few B mesons are produced at the proton energies used by BNB and NuMI. Furthermore, \cite{PhysRevD.100.115039} finds that proton bremsstrahlung processes in which the incident proton radiates an HPS do not improve the SBN Program's sensitivity beyond kaon-decay production channels. Both of these production modes are thus neglected.

\subsubsection{Decay}

\begin{figure}[t]
    \centering
    \includegraphics[width=0.5\textwidth]{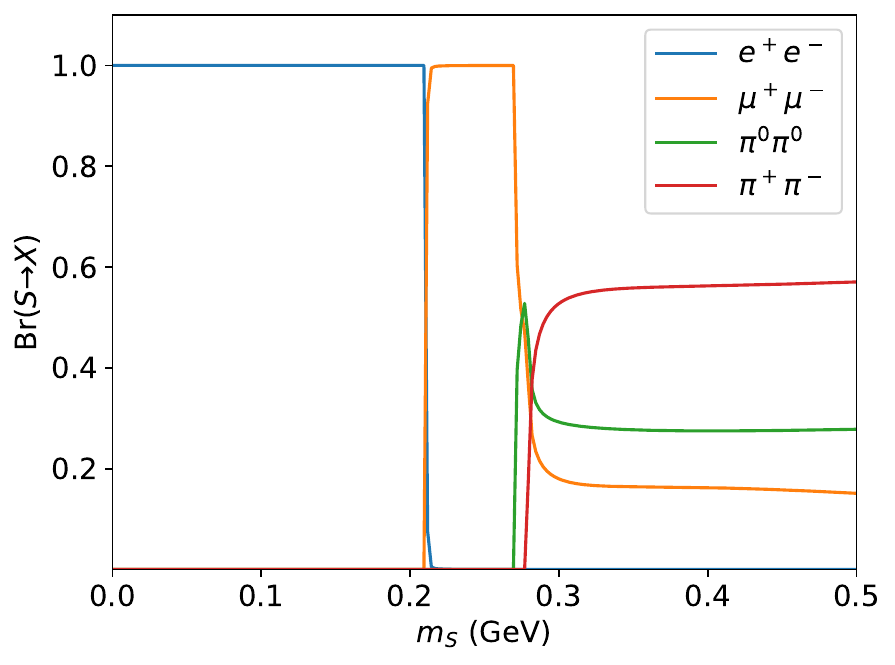}
    \caption{HPS branching ratios.}   
    \label{fig:hps_br}
\end{figure}

For the scalar mass ranges that the SBN Program is sensitive to, HPSs may decay into lepton or pion pairs, as illustrated in the right panel  of \cref{fig:hpsFeynmanDiagrams}. The partial decay width into leptons is given by 
\begin{equation}
    \Gamma(S \rightarrow l^+l^-) = \theta_S^2 \frac{m_l^2 m_S}{8\pi v^2}\left(1-\frac{4m_l^2}{m_S^2}\right)^{3/2}.
\label{HiggsToLeptonsBr}
\end{equation} 
Scalar decay to pions is calculated with chiral perturbation theory, resulting in
\begin{equation}
    \Gamma(S \rightarrow \pi^0\pi^0) = \theta_S^2 \frac{|G_\pi (m_S^2) |^2}{32 \pi v^2 m_S}\left(1 - \frac{4m_\pi^2}{m_S^2}\right)^{1/2},
\label{HiggsBr}
\end{equation} 
with form factor $G_\pi(x) = \frac{2}{9}x + \frac{11}{9}m_\pi^2$ \cite{PhysRevD.100.115039}. $\Gamma(S \rightarrow \pi^+\pi^-) = 2 \times \Gamma(S \rightarrow \pi^0\pi^0)$, since the charged pions comprise two components of the pions' isospin triplet. The branching ratios for the HPSs are shown in \cref{fig:hps_br}.

\subsubsection{Special considerations for kaon kinematics at ICARUS}

Kaons produced in the NuMI target will generally decay in flight in the decay pipe, shortly after being produced.  Portal particles produced from kaon decay in flight (KDIF), and arriving at ICARUS, thus have momenta closely oriented along the direction of the vector pointing from the NuMI target to ICARUS. When those portal particles decay in ICARUS, the summed momentum of their decay products should also be oriented along this direction. Furthermore, the decay products will be highly columnated, as they come from portal particles boosted by the parent kaon's lab frame velocity. 

On the other hand, roughly $13\%$ of protons impinging on the NuMI target punch through and strike the absorber~\cite{NuMIbeam} located roughly 700 m downstream of the target. Kaons produced in proton-absorber interactions typically stop before decaying, leading to kaon decay at rest (KDAR). Scalars arriving at ICARUS that are produced from KDAR are monoenergetic, tend not to have decay products that are as columnated as those coming from KDIF, and have momenta oriented along the direction pointing from the absorber to ICARUS.

Because kaon kinematics can lead to drastically different portal particle signatures in ICARUS, it can be useful to design separate searches for scalars originating from KDIF near the target and those originating from KDAR in the absorber.
Accordingly, settings in the \lstinline{Kaon2HiggsFlux} tool  allow for event filtering based on kaon momentum at time of decay 
and the kaon decay position along the NuMI beam $z$ coordinate.

\subsection{Heavy Neutral Lepton}
\label{sec:moddels_hnl}
First proposed in the late 1970s and early 1980s \cite{glashow1980future,langacker1981grand}, HNLs are fermionic extensions to the SM that mix with the three weakly charged neutrinos via an extension to the PMNS mixing matrix: $U_{\alpha4},\ \alpha= e,\mu,\tau$. The interaction Lagrangian for the HNL field $N$ comes from the weak interactions of the flavored neutrinos:
\begin{equation}
    \mathcal{L}_\text{int} = -\frac{g}{2 \cos\theta_W} \, Z_\mu \, \sum_\alpha U_{\alpha 4} \, \overline{\nu}_\alpha \gamma^\mu P_L N  - \frac{g}{\sqrt{2}} \, W^+_\mu \, \sum_\alpha U_{\alpha 4} \,\overline{\ell_\alpha}  \gamma^\mu P_L N + \text{h.c.},
\end{equation}
where $\theta_W$ is the Weinberg angle and $\text{h.c.}$ denotes the Hermitian conjugate of this expression.
Some HNL models can explain the origin of the masses of the three SM neutrinos via the see-saw mechanism \cite{gell2010complex, yanagida1979proc,mohapatra1980neutrino}. Searches from the MicroBooNE collaboration have demonstrated sensitivity to the $U_{\mu4}$ mixing parameter in the mass range of 10s to 100s of MeV using various channels~\cite{MicroBooNE:2023eef,MicroBooNE:2022ctm}, placing leading exclusion contours for HNL masses in the range $~30-180$ MeV \cite{MicroBooNE:2023eef}. ICARUS and SBND are projected to further probe the $U_{\mu4}$ and $U_{\tau4}$ parameter spaces past existing bounds ~\cite{hnl_decay2, HNL_proj_icarus_numi_2025}. 

\begin{figure}
    \centering
    \begin{subfigure}{0.2\textwidth}
        \centering
        \includegraphics[width=\textwidth]{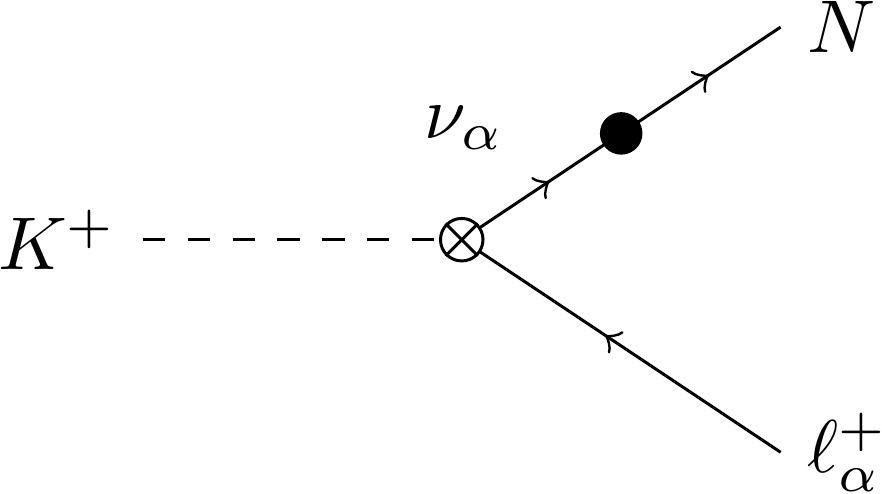}
    \end{subfigure}
    \hspace{0.5cm}
    \begin{subfigure}{0.2\textwidth}
        \centering
        \includegraphics[width=\textwidth]{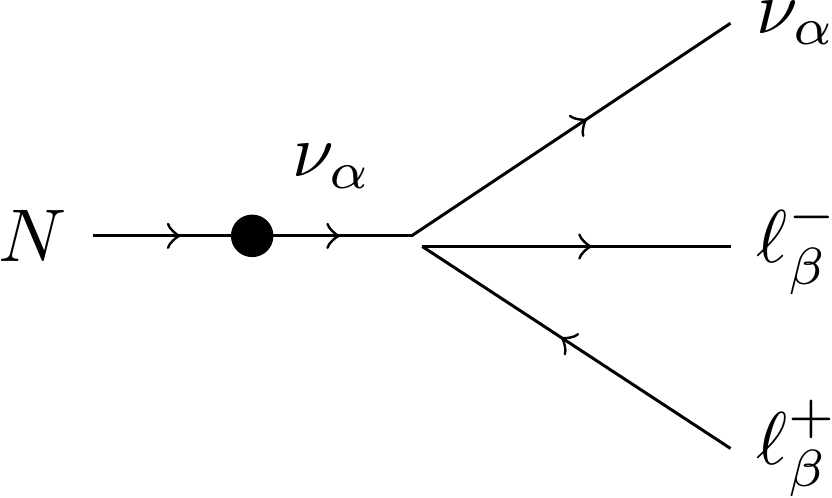}
    \end{subfigure}
    \hspace{0.5cm}
    \begin{subfigure}{0.2\textwidth}
        \centering
        \includegraphics[width=\textwidth]{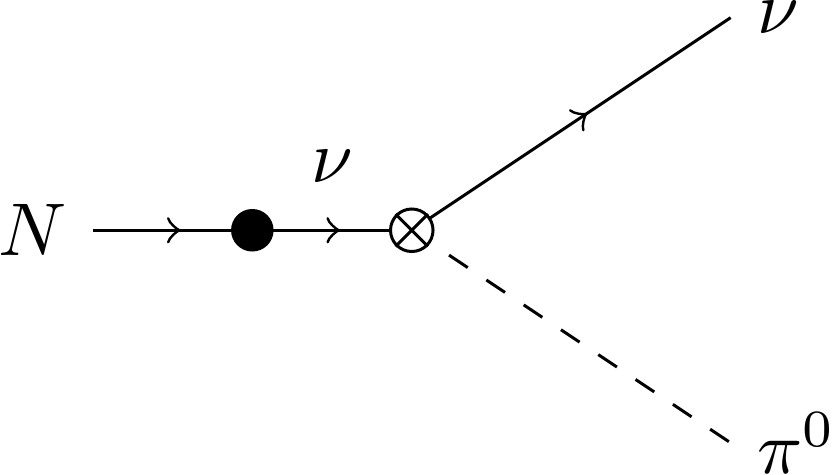}
    \end{subfigure}
    \hspace{0.5cm}
    \begin{subfigure}{0.2\textwidth}
        \centering
        \includegraphics[width=\textwidth]{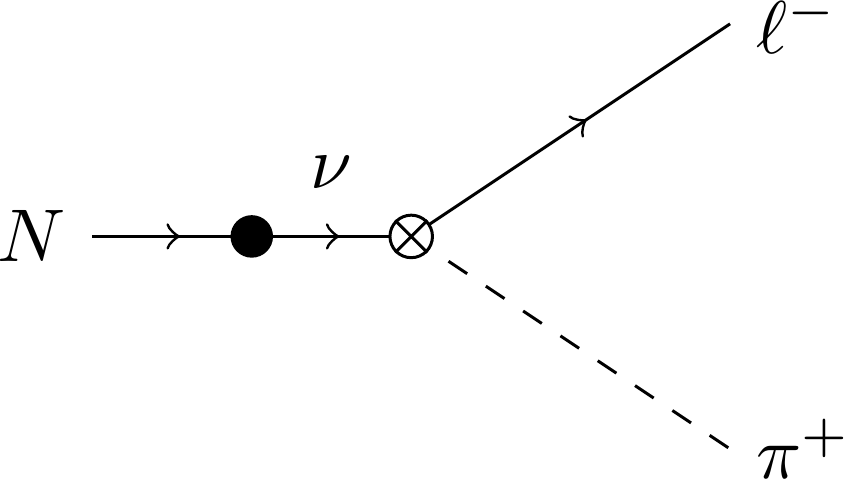}
    \end{subfigure}
    \caption{Feynman diagrams for HNL ($N$) production from kaon decay and the following HNL decay modes: $N\rightarrow \nu_\alpha l^+_\beta l^-_\beta$, $N\rightarrow \nu\pi^0$, and $N\rightarrow \ell_\alpha\pi$. Solid dots represent mixing between HNLs and a SM neutrino $\nu$ (or $\nu_\alpha$, when the neutrino flavor matters), crossed dots represent a weak vertex in the chiral effective theory, and a gapped four-fermion meeting is a (elementary) four-fermion operator.}
    \label{fig:hnl_feynman_diagrams}
\end{figure}

Below, the production and decay of HNLs in \lstinline{MeVPrtl} is described. HNL production via kaon decay is depicted in \cref{fig:hnl_feynman_diagrams}, along with a few HNL decay channels. Specific details on how to generate HNL events with \lstinline{MeVPrtl} are presented in \cref{apn:runHNL}.

\subsubsection{Production}

Kaon decays constitute the main production channel for electron- and muon-coupled HNLs over the parameter space to which the SBN Program is most sensitive. Pions contribute to HNLs with masses below the $\pi$-$e$ and $\pi$-$\mu$ mass differences for $U_{e N}$ and $U_{\mu N}$, respectively, while kaons contribute to HNLs with masses below the $K$-$e$ and $K$-$\mu$ mass differences for $U_{e N}$ and $U_{\mu N}$, respectively.
The branching ratio of a meson $M$ decaying into a lepton and HNL pair, $\ell_{\alpha}N$, is given by \cite{berryman2020searches}
\begin{equation}
    \operatorname{Br}\left(M^{+} \rightarrow \ell_\alpha^{+} N\right)=\operatorname{Br}\left(M^{+} \rightarrow \ell_\alpha^{+} \nu\right)\left(\frac{\left|U_{\alpha N}\right|^2}{1-\left|U_{\alpha N}\right|^2}\right) \rho_N\left(\frac{m_{\ell_\alpha}^2}{m_{M}^2}, \frac{m_N^2}{m_{M}^2}\right),
\end{equation}
with $\rho_N$ a function of the lepton, HNL, and meson masses that takes into account helicity unsuppression and phase space differences for a massive neutrino: 
\begin{equation}
\rho_N(x, y)=\frac{\left(x+y-(x-y)^2\right) \sqrt{1+x^2+y^2-2(x+y+x y)}}{x(1-x)^2} .
\end{equation}
Three-body kaon decays have a minor contribution to the HNL flux and can be safely neglected for most HNL masses.

The decay of the mesons into HNLs is polarized due to the parity violation of the weak force.  The decay of the HNLs must therefore be computed using polarized HNLs, rather than being helicity averaged with even weighting for each helicity.  As shown in ref.\ \cite{HNLAn1}, the decay of the polarized HNLs can be written in terms of the average polarization of the HNLs produced in the meson decays.  This average polarization, in the rest frame of the parent meson, is computed as \cite{HNLAn1}
\begin{equation}\label{eq:pol-prod}
    P = \frac{(y_N^2-y_{\ell_\alpha}^2) \sqrt{(1-y_{\ell_\alpha}^2)^2-2y_N^2(1+y_{\ell_\alpha}^2) + y_N^4}}{(1 - y_{\ell_\alpha}^2)y_{\ell_\alpha}^2 + 2 y_{\ell_\alpha}^2y_N^2 + (1-y_N^2)y_N^2},
\end{equation} 
where $y_N = m_N/m_M$, $y_{\ell_\alpha} = m_{\ell_\alpha}/m_M$, and $m_M$ is the mass of the parent meson. The polarization in the CP-conjugate decay has the opposite sign.

In the NuMI beamline, HNLs can also originate from decay of the $\tau$ lepton. In the flavor basis, this production mechanism is mediated through $\nu_\tau$/HNL mixing via $U_{\tau N}$, where $\nu_\tau$ can be thought of as an intermediate decay product. A full accounting for all the ways in which $\tau$ leptons can produce a $\tau$-coupled HNL is not available in \lstinline{MeVPrtl} at present.  

\subsubsection{Decay}

For kaon-produced HNLs, the HNL can decay in three channels that are accessible to the SBN Program.  These are the same-flavor leptonic channel $N \to \nu_\alpha \, \ell^+_\beta \, \ell^-_\beta$ and two channels with pions, $N \to \nu_\alpha \, \pi^0$ and $N \to \ell_\alpha^- \, \pi^+$.  The widths for decays in these channels are given by \cite{hnl_decay2,hnl_decay1}
\begin{enumerate} [(a)]
   \item \textit{$N\rightarrow \nu_\alpha l^+_\beta l^-_\beta$:}

   \begin{equation}
\begin{aligned}
& \Gamma\left(N \rightarrow \nu_\alpha l^{+}_\beta l^{-}_\beta\right)= \frac{G_{\mathrm{F}}^2 m_N^5}{192 \pi^3}\left|U_{\alpha 4}\right|^2\left[\left(g_L g_R+\delta_{\alpha \beta} g_R\right) I_1\left(0, \frac{m_{l_\beta}}{m_N}, \frac{m_{l_\beta}}{m_N}\right)\right. \\
&\left.+\left(g_L^2+g_R^2+\delta_{\alpha \beta}\left(1+2 g_L\right)\right) I_2\left(0, \frac{m_{l_\beta}}{m_N}, \frac{m_{l_\beta}}{m_N}\right)\right],
\end{aligned}
\end{equation}
with $\delta_{\alpha\beta}$ a Kronecker delta symbol, $G_{\mathrm{F}}$ the Fermi constant, $g_L=-1/2+\sin^2\theta_W$, $g_R=\sin^2\theta_W$ and $I_{1,2}$ given by:
\begin{equation}
\begin{aligned}
& I_1(x, y, z)=12 \int_{(x+y)^2}^{(1-z)^2} \frac{d s}{s}\left(s-x^2-y^2\right)\left(1+z^2-s\right) \sqrt{\lambda\left(s, x^2, y^2\right)} \sqrt{\lambda\left(1, s, z^2\right)}, \\
& I_2(x, y, z)=24 y z \int_{(y+z)^2}^{(1-x)^2} \frac{d s}{s}\left(1+x^2-s\right) \sqrt{\lambda\left(s, y^2, z^2\right)} \sqrt{\lambda\left(1, s, x^2\right)}, \\
& \lambda(a, b, c)=a^2+b^2+c^2-2 a b-2 b c-2 c a .
\end{aligned}
\end{equation}

    \item \textit{$N\rightarrow \nu\pi^0$}: 
    
    \begin{equation}
    \Gamma\left(N \rightarrow \nu \pi^0\right)=\sum_\alpha \frac{G_{\mathrm{F}}^2 f_\pi^2 m_N^3\left|U_{\alpha 4}\right|^2}{32 \pi}\left[1-\left(\frac{m_\pi}{m_N}\right)^2\right]^2.
    \end{equation}
    where $f_\pi$ is the pion decay constant.

    \item \textit{$N\rightarrow \ell_\alpha\pi$:}
    
    \begin{equation}
    \Gamma\left(N \rightarrow \ell_\alpha^{ \pm} \pi^{\mp}\right)=\left|U_{\alpha 4}\right|^2 \frac{G_{\mathrm{F}}^2 f_\pi^2\left|V_{u d}\right|^2 m_N^3}{16 \pi} I\left(\frac{m_{\ell_\alpha}^2}{m_N^2}, \frac{m_\pi^2}{m_N^2}\right),
    \end{equation}

    with
    \begin{equation}
     I(x, y)= \left [ \left(1+x-y\right)\left(1+x\right)-4x\right ] \sqrt{\lambda\left(1, x, y\right)}.
    \end{equation}
  
\end{enumerate}

Other channels either cannot be measured (such as $\nu\nu\nu$) or have a significantly lower branching ratio. Nevertheless, their decay widths are also computed to calculate the total decay width, and are given by

\begin{enumerate} [(a)]
    \item \textit{$N \rightarrow \nu\nu\nu$:}

    \begin{equation}
     \Gamma\left(N \rightarrow \nu\nu\nu \right)= \sum_\alpha  \left | C_{4 \alpha} \right |^2 \frac{G_{\mathrm{F}}^2 m_N^5}{192 \pi^3},
    \end{equation}
    with 
    \begin{equation}
    \sum_\alpha  \left | C_{4 \alpha} \right |^2 = \sum_{\alpha, \beta, j} U_{ \alpha 4}^{*} U_{ \alpha j}^{} U_{ \beta 4}^{}U_{ \beta j}^{*} \approx \sum_\alpha  \left | U_{4 \alpha} \right |^2.
    \end{equation}

    \item \textit{$N\rightarrow l_\alpha^-l_\beta^+\nu_\beta$, $\alpha \neq \beta$:}

    \begin{equation}
 \Gamma\left(N\rightarrow l_\alpha^-l_\beta^+\nu_\beta \right)=\frac{G_{\mathrm{F}}^2 m_N^5}{192 \pi^3}\left|U_{\alpha 4}\right|^2 I_1\left(\frac{m_{l_\alpha}}{m_N}, 0, \frac{m_{l_\beta}}{m_N} \right). 
    \end{equation}

\end{enumerate}

\begin{figure}[h]
    \centering
    \begin{subfigure}{0.45\textwidth}
        \centering
        \includegraphics[width=\textwidth]{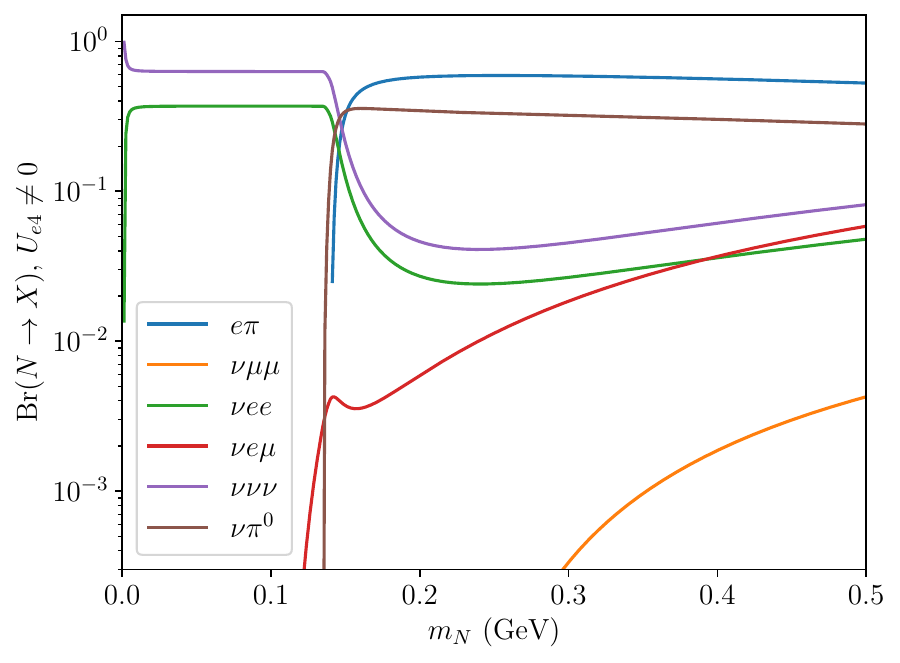}
    \end{subfigure}
    \begin{subfigure}{0.45\textwidth}
        \centering
        \includegraphics[width=\textwidth]{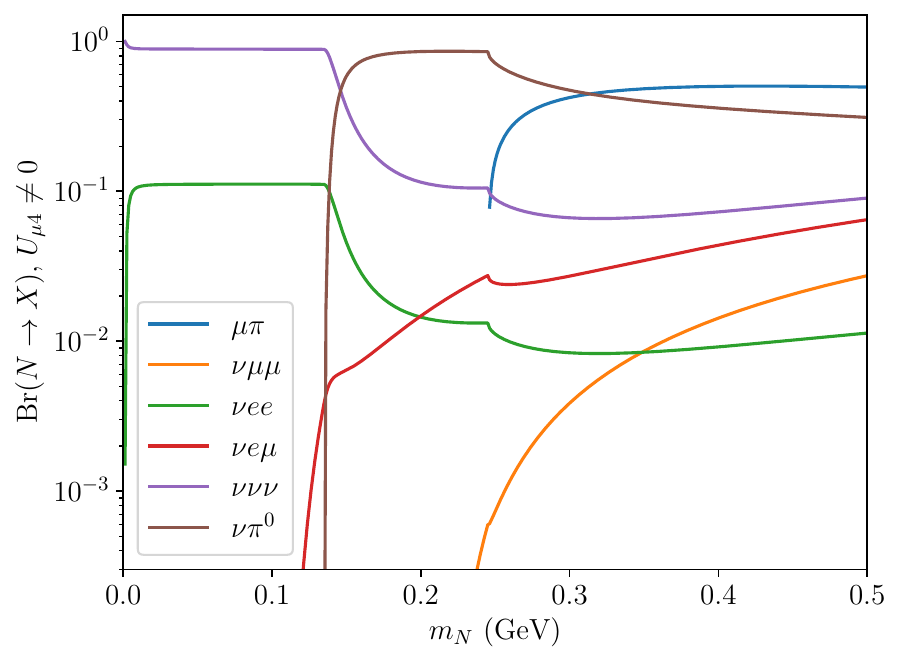}
    \end{subfigure}
    \begin{subfigure}{0.45\textwidth}
        \centering
        \includegraphics[width=\textwidth]{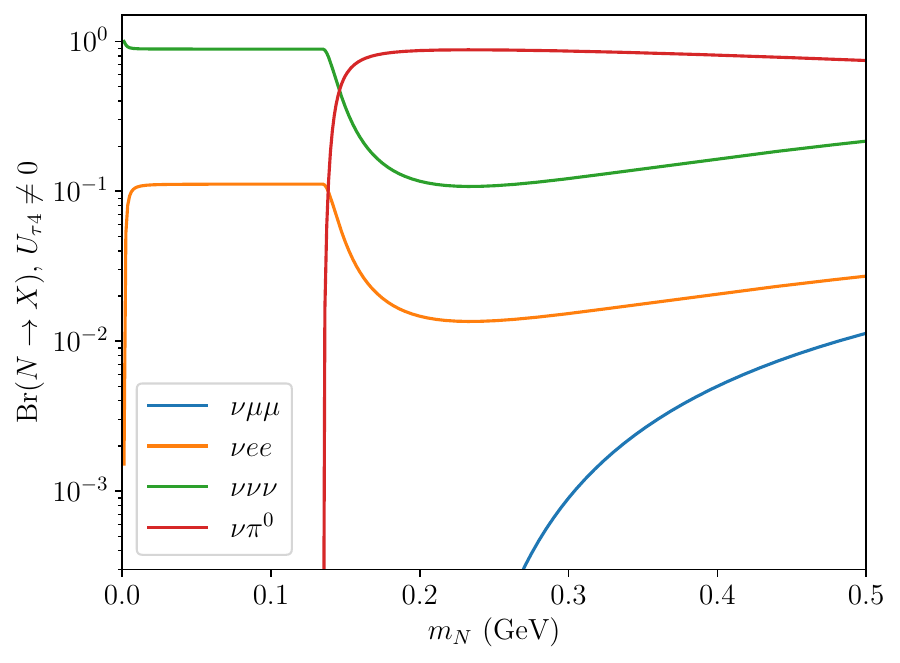}
    \end{subfigure}
    \begin{subfigure}{0.45\textwidth}
        \centering
        \includegraphics[width=\textwidth]{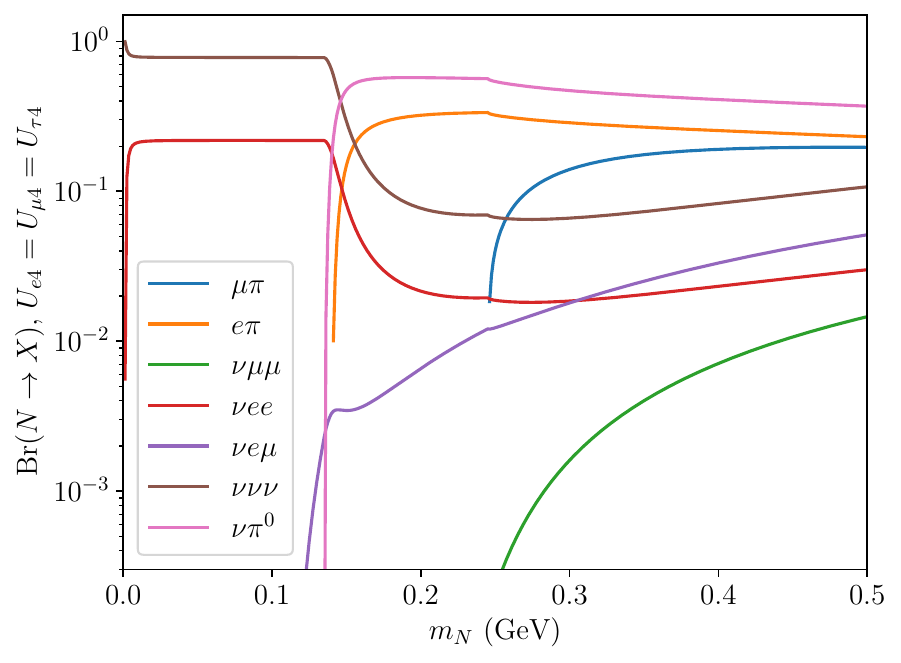}
    \end{subfigure}
    
    \caption{Branching ratios of HNL decays to different final states for electron only (top-left), muon only (top-right), tau only (bottom-left), and mixed (bottom-right) couplings as computed by \lstinline{MeVPrtl}.}   
    \label{fig:HNLdecaywvalidationemu}
\end{figure}

All these widths are valid for Dirac HNLs. In case of the Majorana HNLs the expressions are multiplied by a factor of two to account for the new possible charge-conjugated final states. The corresponding branching ratios are not affected by the Majorana nature of the HNL. The branching ratios of HNL decays to different final states for various couplings, as computed by \lstinline{MeVPrtl}, are shown in \cref{fig:HNLdecaywvalidationemu}.

\subsubsection{Decay kinematics: 2 and 3-body decay}

The full polarization-dependent kinematics of the HNL decays are taken into account to produce a precise simulation of the process.

In the two-body decay mode $N \to \ell_\alpha \pi$, the differential width in the HNL rest frame is~\cite{Ballett:2019bgd}
\begin{equation}
\frac{d\Gamma(N\to \ell_\alpha^- \pi^+)}{d\cos\theta} = |U_{\alpha 4}|^2 |V_{ud}|^2 \frac{G_F^2 f_\pi^2 m_N^3}{8} \left[\frac{1+P}{2} I_1^+(\xi_\alpha^2,\xi_\pi^2;\theta) + \frac{1-P}{2} I_1^-(\xi_\alpha^2,\xi_\pi^2;\theta)\right],
\end{equation}
where
\begin{equation}
    I_1^{\pm}(x,y;\theta) = \frac{1}{4\pi} \lambda^{1/2}(1,x,y) \left[(1-x)^2 - y (1+x)  \pm (x-1) \lambda^{1/2}(1,x,y) \cos\theta\right],
\end{equation}
$\xi_i = m_i / m_N$, $V_{ud}$ is a CKM matrix element, and $\theta$ is the azimuthal angle of the outgoing lepton.  The CP conjugate decay has $I_1^\pm \to I_1^\mp$.  The polarization of the $N \to \nu_\alpha \pi^0$ decays is not simulated as the neutrino escapes the detector, precluding a measurement of the angular distribution.

In the three-body scenario $N \rightarrow \nu l^{+}_\alpha l^{-}_\alpha $ both the non-trivial kinematics induced by the polarization of the HNL (relevant for Dirac HNLs) and the electroweak contribution to the vertex are considered. The differential decay width in the HNL reference frame is given by \cite{HNLAn1,HNLAn2}
\begin{equation}
    \frac{d\Gamma\left(N \rightarrow \nu l_{\alpha}^{+} l_{\alpha}^{-}\right)}{dm_{ll}^2d\cos\theta_{ll} dm_{\nu m}^2d\gamma_{ll}d\phi} = \frac{1}{(2\pi)^5}\frac{1}{64 m_N^3}\sum_{j=1}^{13}C_{j\alpha\beta}K_j.
\end{equation}  
The differential width is used to simulate angular distributions.
Here, the nonzero $C_{j\alpha\beta}$ coefficients are given in \cref{tab:HNLAnCiCoef} in the case of the HNL only interacting with the SM by mixing with the light neutrinos. 
The definitions $g_{L,\alpha\beta} = \left|U_{\beta 4}\right| \left[ \delta_{\alpha\beta} - \frac{1}{2} \left( 1 - 2\sin^2\theta_W\right)\right]$ and $g_{R,\alpha\beta} = \delta_{\alpha\beta} |U_{\beta 4}| \sin^2\theta_W$ are used.
The Lorentz-invariant quantities $K_j$ are defined as 

\begin{equation}
\begin{aligned}
& K_1 = \frac{1}{2}m_n m_p \left( m_N^2 - m_{ll}^2 \right), \\
& K_4 = \frac{1}{4} \left( m_{\nu m}^2 -m^2_m \right) \left( m_{N}^2 + m^2_m - m_{\nu m}^2 \right),  \\
& K_5 = \frac{1}{4} \left( m_{\nu m}^2 + m_{ll}^2 -m^2_p \right) \left( m_{N}^2 + m^2_m - m_{ll}^2 - m_{\nu m}^2 \right),  \\
& K_8 = \frac{P}{2} m_n m_p \left( m_{N}^2 - m_{ll}^2\right) \cos \theta_{ll},  \\
& K_9 = \frac{P m_N}{2} \left( m_{\nu m}^2 - m_{m}^2\right) \left( |p_m| (\cos\gamma_{ll}\sin\theta_{ll}\sin\theta_{\nu m} - \cos \theta_{ll} \cos\theta_{\nu m}) - \frac{m_N^2-m_{ll}^2}{2m_N}\cos\theta_{ll} \right),  \\
& K_{10} = -\frac{P m_N}{2} \left( m_N^2 + m_n^2 -m_{\nu m}^2 - m_{ll}^2\right) |p_m| \left(\cos\gamma_{ll}\sin\theta_{ll}\sin\theta_{\nu m} - \cos \theta_{ll} \cos\theta_{\nu m}\right),  \\
\end{aligned}
\end{equation}
where $P$ is the polarization of the HNL given by \cref{eq:pol-prod}, $m_n$ and $m_p$ the masses of the negatively and positively charged leptons respectively, $m_{ll}$ the invariant mass of the charged lepton pair, $m_{\nu m}$ the invariant mass of the neutrino-negatively charged lepton system, $\theta_{ll}$ the angle between the HNL spin direction and the sum of the charged lepton momenta $p_{\ell\ell}$, $\theta_{\nu m}$ is the angle between the outgoing neutrino and negatively charged lepton, $\gamma_{ll}$ the angle of rotation of the charged lepton subsystem around the direction of $p_{ll}$, and $p_m$ is the magnitude of the three momentum of the negatively charged lepton. The full expression for the last of these can be found in terms of our kinematic variables in Ref.\ \cite{HNLAn2}.  The CP conjugate decay has the opposite sign on the terms proportional to $P$.

\begin{table} [h]\centering
\begin{tabular}{|l|lll|}
\hline
$C_{j\alpha\beta}$ & Dirac N & Dirac $\bar{N}$ & Majorana N \\ \hline
$C_1$     &  $g_{L,\alpha\beta}g_{R,\alpha\beta}$  &    $g_{L,\alpha\beta}g_{R,\alpha\beta}$     &  $2 g_{L,\alpha\beta}g_{R,\alpha\beta}$          \\
$C_4$     &  $g_{L,\alpha\beta}^2$       &         $g_{R,\alpha\beta}^2$                        &        $g_{L,\alpha\beta}^2+ g_{R,\alpha\beta}^2$    \\
$C_5$     &   $g_{R,\alpha\beta}^2$      &           $g_{L,\alpha\beta}^2$                     &       $g_{L,\alpha\beta}^2+ g_{R,\alpha\beta}^2$     \\ \hline
$C_8$     &  $-g_{L,\alpha\beta}g_{R,\alpha\beta}$       &         $g_{L,\alpha\beta}g_{R,\alpha\beta}$                          &   0         \\
$C_9$     &  $-g_{L,\alpha\beta}^2$      &          $g_{R,\alpha\beta}^2$                      &   $g_{R,\alpha\beta}^2-g_{L,\alpha\beta}^2$         \\ $C_{10}$    &  $-g_{R,\alpha\beta}^2$       &            $g_{R,\alpha\beta}^2$                      &  $g_{R,\alpha\beta}^2-g_{L,\alpha\beta}^2$          \\ \hline
\end{tabular}
\caption{Coefficients determining angular anisotropy in HNL decay.  The common factor of $64 G_F^2$ is factored out of these expressions.}
\label{tab:HNLAnCiCoef}
\end{table}

\subsection{Heavy QCD Axion}
\label{sec:moddels_axion}
The final dark sector model presently incorporated into \lstinline{MeVPrtl} is that of heavy ($m_a \sim \text{GeV}$) QCD axions \cite{Dimopoulos:1979pp,Tye:1981zy,Holdom:1982ex,Rubakov:1997vp,Berezhiani:2000gh,Gherghetta:2016fhp,Dimopoulos:2016lvn,Fukuda:2017ywn,Agrawal:2017ksf,Lillard:2018fdt,Gaillard:2018xgk,Hook:2019qoh,Gherghetta:2020keg,Gherghetta:2020ofz,Co:2022aav,AxionSBN,Dunsky:2023ucb}, which have a gluonic coupling to the SM by definition. The existence of such a particle is motivated as a solution to the strong CP problem~\cite{Belavin:1975fg,tHooft:1976rip,Jackiw:1976pf,Callan:1976je,Crewther:1979pi}. For the mass range of interest, the axion is unstable and is not a dark matter particle candidate. On the other hand, the low energy scale of the axion decay constant for this model, of order $\text{PeV}$, would alleviate the axion quality problem~\cite{Kamionkowski:1992mf,Barr:1992qq,Ghigna:1992iv,Holman:1992us} otherwise present in many axion dark matter models. Many searches have been proposed and completed to probe parts of ALP parameter space~\cite{Essig:2010gu,Dolan:2017osp,Harland-Lang:2019zur,Dobrich:2019dxc,Dent:2019ueq,Brdar:2020dpr,AxionArgoNEUT,AxionDUNE,Jang:2022tsp,Yu:2025mlx}. The SBN Program is sensitive to two distinct production modes of ALPs: kaon decay and mixing with pseudoscalar mesons $\pi^0$, $\eta$, and $\eta'$. Once produced, axions could decay to a variety of final states: $\gamma\gamma,\  \pi\pi\pi,\  \pi\pi\gamma$, as well as $\mu^+\mu^-$ and $e^+e^-$ if the axion has leptonic couplings. These decays are considered in \lstinline{MeVPrtl}. 

The model consists of an axion $a$ with couplings $c_1$, $c_2$, $c_3$ to the SM $U(1)_Y$, $SU(2)_L$, and $SU(3)_C$ gauge bosons, respectively, as well as leptonic couplings that can be included at tree-level or induced by the other couplings at loop level. The axion mass $m_a$ and decay constant $f_a$ are additional free parameters, though $f_a \gg \text{TeV}$ is required to keep the mixing between the axion and SM particles small. The interaction Lagrangian is described by

\begin{equation}
    \mathcal{L}_\text{int} =  \mathcal{L}_\text{gauge} + \mathcal{L}_\text{lepton},
\end{equation}
with 
\begin{equation}
    \mathcal{L}_\text{gauge} = c_3 \frac{\alpha_3}{8\pi f_a}aG\tilde{G} + c_2 \frac{\alpha_2}{8\pi f_a}aW\tilde{W} + c_1 \frac{\alpha_1}{8\pi f_a}aB\tilde{B}
\end{equation}
and 
\begin{equation}
\mathcal{L}_\text{lepton} = \sum_{\ell=e,\mu,\tau} c_l\frac{\partial_\mu a}{2 f_a}\bar{\ell}\gamma^\mu \gamma_5 \ell,
\label{eqn:axionLepton}
\end{equation}
where $\alpha_1,\alpha_2,\alpha_3$ are the SM gauge couplings for $U(1)_Y$, $SU(2)_L$, and $SU(3)_C$ respectively.

\lstinline{MeVPrtl} defaults to the ``co-dominance'' scenario in which all the gauge couplings are equal ($c_1=c_2=c_3$), and the assumption that $c_\mu$ is the only non-zero leptonic coupling. \Cref{apn:runALP} provides specific details on how to generate ALP events with \lstinline{MeVPrtl}.

\subsubsection{Production}
\label{sec:models_axion_production}

\begin{figure}[t]
    \centering
    \begin{subfigure}{0.35\textwidth}
        \centering        \includegraphics[width=\textwidth]{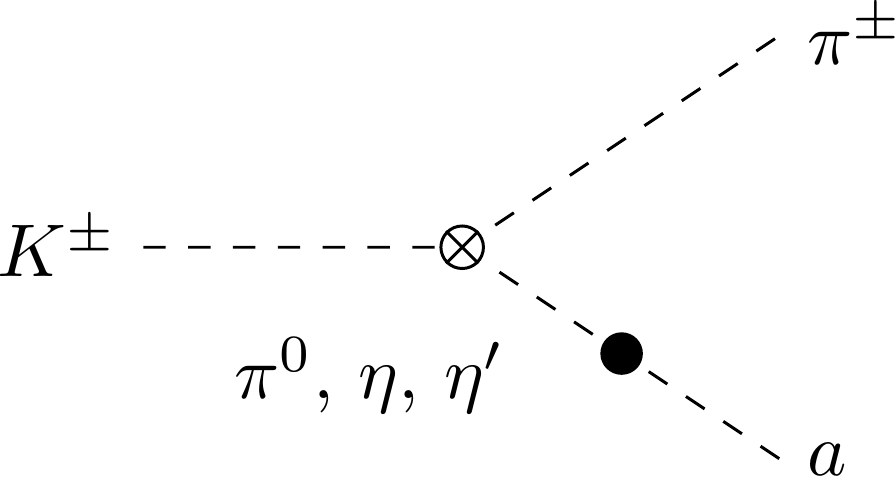}
    \end{subfigure}
    \hspace{1cm}
    \begin{subfigure}{0.4\textwidth}
        \centering        \includegraphics[width=\textwidth]{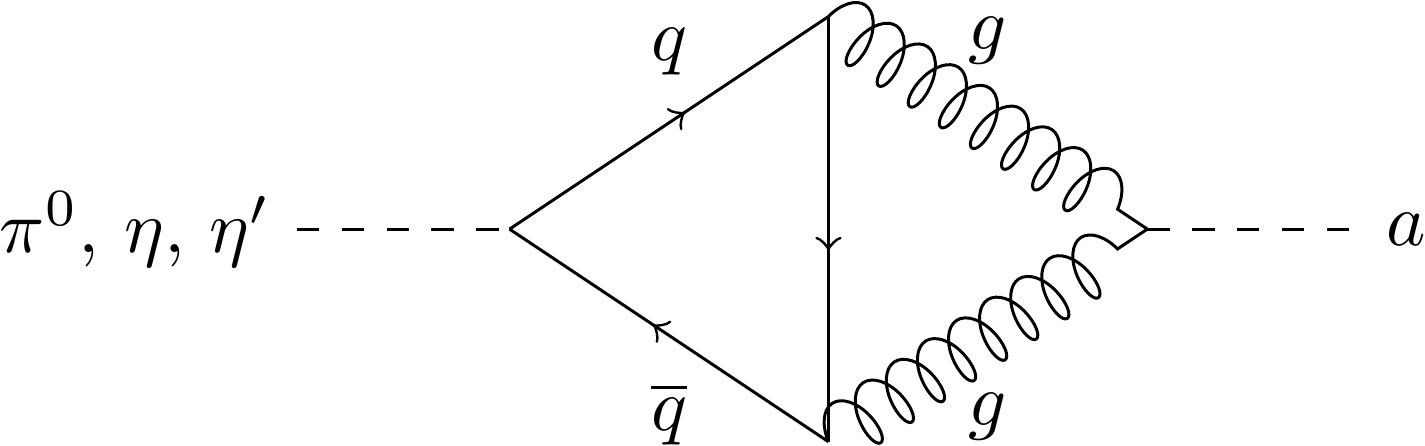}
    \end{subfigure}
    \caption{Feynman diagrams for ALP production via kaon decay (left) and mixing with pseudoscalars $\pi^0$, $\eta$, or $\eta'$ (right). The crossed dot is a weak vertex in the chiral effective theory. The solid dot on the left is a shorthand representation of the axion mixing with the pseudoscalars, i.e. the process shown in detail on the right. The mixing is mediated by quarks (solid lines) and gluons (looped lines).}   
    \label{fig:axion_prod_feynman_diagrams}
\end{figure}

Axions can be produced at the SBN Program from kaon decay or through mixing with pseudoscalar mesons $\pi^0$, $\eta$, and $\eta'$, as depicted in \cref{fig:axion_prod_feynman_diagrams}. Production via kaon decay dominates below the $m_K - m_\pi$ axion mass threshold \cite{Berger_Putnam_2024}. This production channel is not explicitly generated by \lstinline{MeVPrtl}, given its shared phenomenology with HPSs: to ``simulate'' ALPs originating from kaon decay, users of \lstinline{MeVPrtl} are instead instructed to simulate kinematically identical (i.e.\:same mass) HPS events, and then reweight them as axions according to the production and decay branching fractions presented below. At higher masses, the ALPs are primarily produced via mixing with pseudoscalar mesons; this production mode is explicitly simulated by \lstinline{MeVPrtl} and discussed below under point (b) \textit{Mixing with $\pi^0$, $\eta$, and $\eta'$}. Reference \cite{Brdar:2020dpr} points out that the Primakoff process $\gamma + A \to a + A$ in which a photon scatters off of a nucleus to produce an ALP would be the dominant production mechanism if the ALP's gluonic coupling were subdominant to its photonic coupling. Adding that production mechanism to \lstinline{MeVPrtl} is an interesting avenue of further work.

Production of ALPs in kaon decays is described as follows.  CP violation suppresses the production of axions (which are CP odd) from $K_L^0$, leaving just the following relevant branching fractions for axion production via kaon decay \cite{Bauer_2021, Bauer_2022} 

\begin{equation}
    \text{Br}\left(K^\pm \to a \pi^\pm\right) = \frac{\tau_{K^\pm}}{\tau_{K^0_S}}\frac{2 f_\pi^2 c_3^2}{f_a^2} \left(\frac{m_K^2-m_a^2}{4 m_K^2 - 3 m_a^2 - m_\pi^2}\right)^2 \sqrt{\frac{\lambda(1, m_\pi^2/m_K^2, m_a^2/m_K^2)}{1 - 4 m_\pi^2/m_K^2}} \text{Br}\left(K^0_S \to \pi^+\pi^-\right)\,
\end{equation}
and
\begin{equation}
    \text{Br}(K^0_S \to \pi^0 \, a) \approx \frac{\tau_{K^0_s}}{\tau_{K^\pm}}\text{Br}(K^\pm \to \pi^\pm \, a), 
\label{eqn:Ks_to_a}
\end{equation}
where $\tau_{K}$ are the respective kaon lifetimes and $f_\pi$$\approx 130~\text{MeV}$ is the pion decay constant. Production from $K^0_S$ is highly suppressed by the ratio $\frac{\tau_{K_s}}{\tau_{K^\pm}}$ in \cref{eqn:Ks_to_a}, so to good approximation, only charged kaons can significantly produce the axions. ALP production via kaon decay is not explicitly simulated within \lstinline{MeVPrtl}, but users may reweight generated HPS events coming from charged kaons to the axion model.

ALP production by mixing is implemented by the \lstinline{Meson2ALP} module, which inherits from the \lstinline{MeVPrtlFlux} interface. 

The mixing angles are given by \cite{AxionSBN,AxionDUNE}

\begin{equation}
    \begin{split}
        \theta_{a\pi} &= \frac{1}{6\sqrt{2}}\frac{f_\pi}{f_a}\frac{m_a^2}{m_a^2 - m_\pi^2}, \\
        \theta_{a\eta} &= \frac{1}{2\sqrt{3}}\frac{f_\pi}{f_a}\frac{m_a^2 - 4 m_\pi^2/9}{m_a^2 - m_\eta^2}, \\
        \theta_{a\eta'} &= \frac{1}{2\sqrt{6}}\frac{f_\pi}{f_a}\frac{m_a^2 - 16 m_\pi^2/9}{m_a^2 - m_{\eta'}^2}\, .
    \end{split}
\end{equation}
The rate of mixing with a meson $m$ is then given by $|\theta_{am}|^2$. 

In principle, the mixing of the axion with the meson will change the full kinematics of the interaction that produced it. Simulating that is complicated, as it would require integration into \lstinline{Pythia8} and would depend on the axion mass. Instead, an approximation is employed as in phenomenology papers \cite{AxionDUNE,AxionSBN} and the ArgoNeuT search \cite{AxionArgoNEUT}: the energy of the axion is set to that of the parent meson and the axion momentum is scaled so that it is on-shell while keeping the direction fixed. If the axion mass is greater than the meson energy, the event is dropped.

\subsubsection{Decay}
\label{sec:models_axion_decay}

\begin{figure}
    \centering
    \begin{subfigure}{0.2\textwidth}
        \centering
        \includegraphics[width=\textwidth]{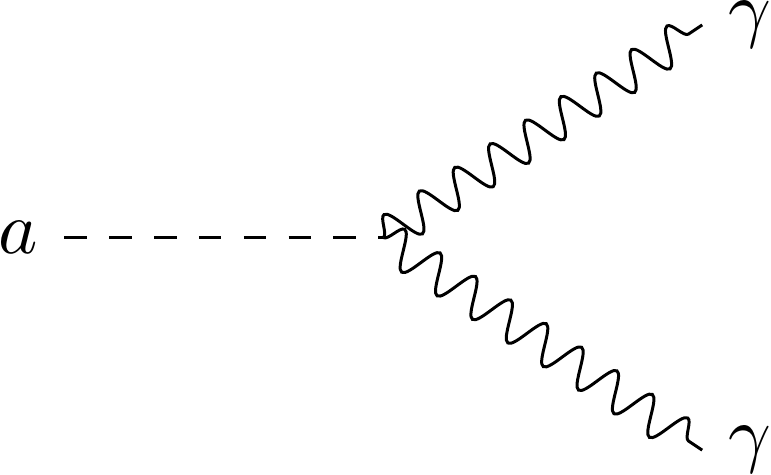}
    \end{subfigure}
    \hspace{0.5cm}
    \begin{subfigure}{0.2\textwidth}
        \centering
        \includegraphics[width=\textwidth]{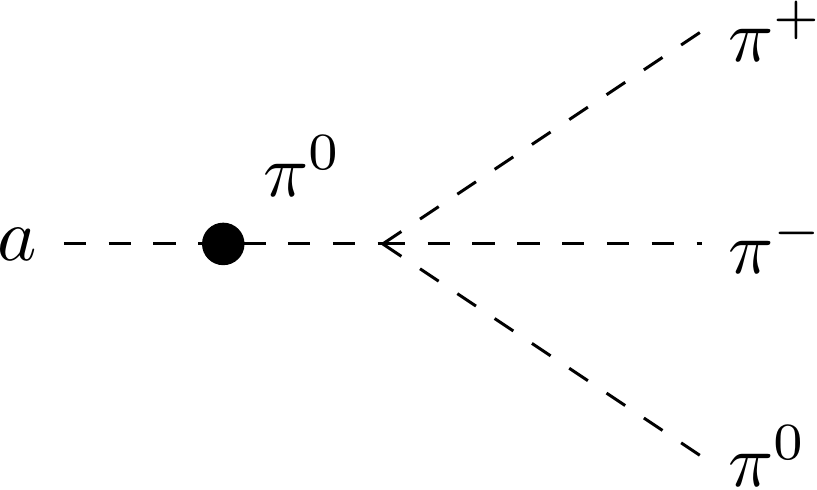}
    \end{subfigure}
    \hspace{0.5cm}
    \begin{subfigure}{0.2\textwidth}
        \centering
        \includegraphics[width=\textwidth]{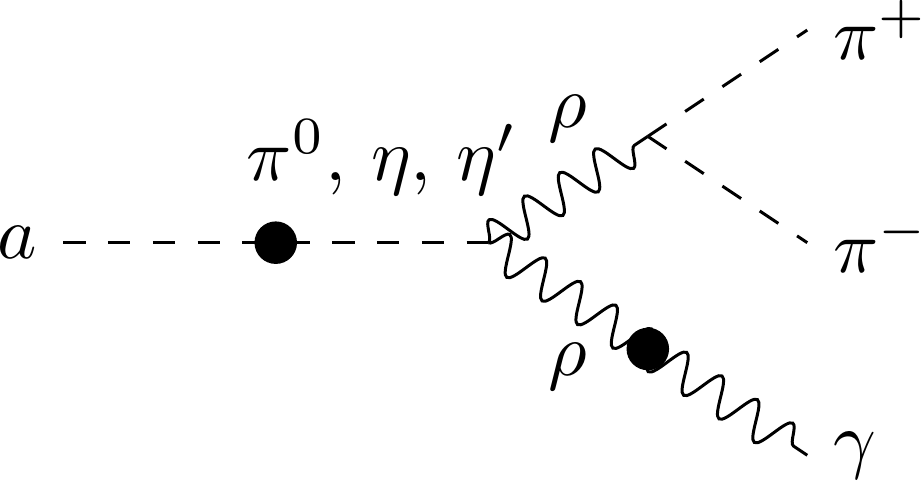}
    \end{subfigure}
    \hspace{0.5cm}
    \begin{subfigure}{0.2\textwidth}
        \centering
        \includegraphics[width=\textwidth]{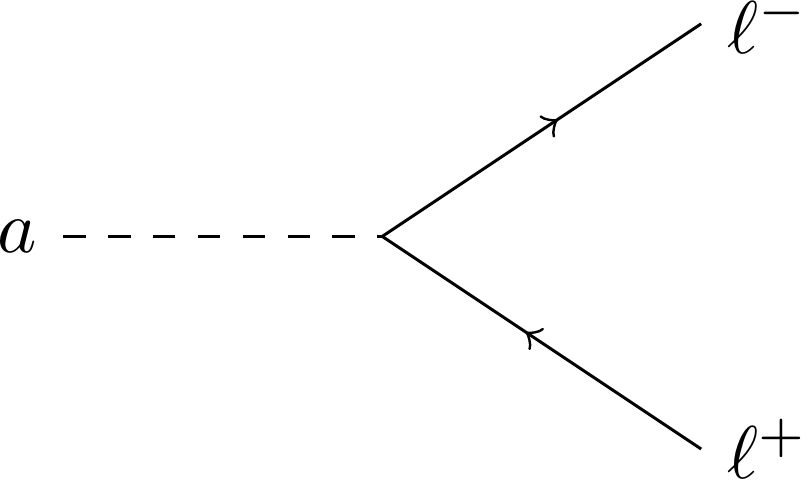}
    \end{subfigure}
    \caption{Feynman diagrams for the following ALP decays: $a\to\gamma\gamma$, $a\to 3\pi$, $a\to 2\pi\gamma$, and $a \to\ell\ell$. The solid dots represent ALP/meson or $\rho$/$\gamma$ mixing.}
    \label{fig:alp_decay_feynman_diagrams}
\end{figure}

\begin{figure}[t]
    \centering    \includegraphics[width=0.5\textwidth]{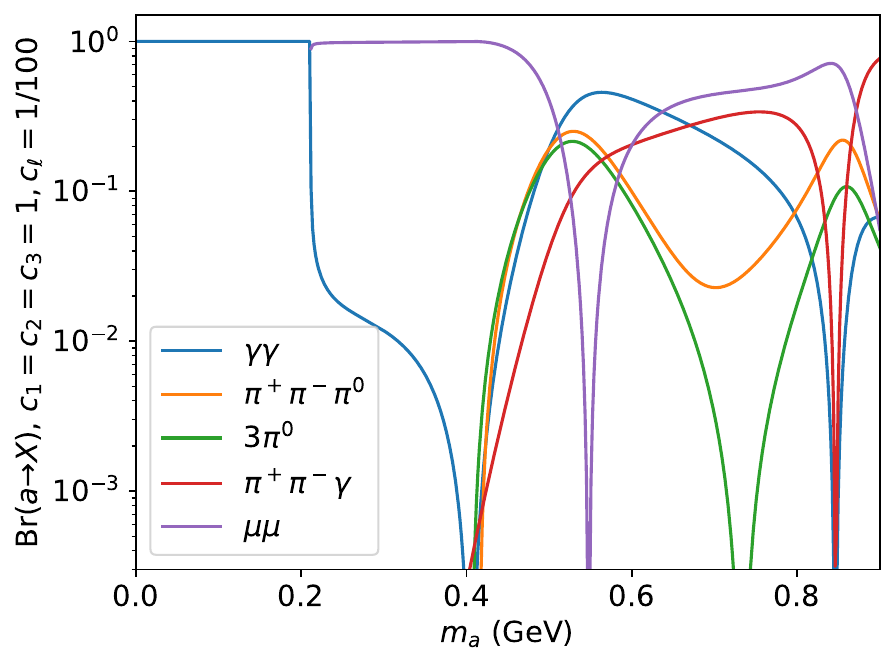}
    \caption{ALP branching fractions for $c_\ell = c_\mu =0.01$ and $c_1 = c_2 = c_3 = 1$. A plot of the corresponding axion decay widths has been verified to be consistent with that shown in \cite{AxionSBN}, but omitted the comparison from this note for brevity.}   
    \label{fig:alp_br}
\end{figure}

The ALP decay widths are enumerated below. Feynman diagrams for some of the processes are shown in \cref{fig:alp_decay_feynman_diagrams}. Fig. \ref{fig:alp_br} shows the branching ratios for ALPs assuming the default coupling strengths in \lstinline{MeVPrtl}: $c_\ell = c_\mu =0.01$ and $c_1 = c_2 = c_3 = 1$.

\begin{enumerate}[(a)]
    \item \textit{Photons}: 
    
    The width is given by \cite{AxionSBN,AxionArgoNEUT,AxionDUNE}
    \begin{equation}
        \Gamma_{a\to\gamma\gamma} = \frac{\alpha^2|c_\gamma|^2m_a^3}{256 \pi^3 f_a^2}\,,
    \end{equation}
    where $\alpha$ is the fine structure constant and $c_\gamma$ is the axion coupling to photons, which gains contributions from all axion couplings once loop diagrams are included \cite{GrillidiCortona:2015jxo,AxionSBN}:
    \begin{multline}
    \label{eqn:cgamma}
        c_\gamma = c_3\left( -1.93 + \frac{1}{3}\frac{m_a^2}{m_a^2-m_\pi^2} + \frac{8}{9}\frac{m_a^2 - 4m_\pi^2/9}{m_a^2 - m_\eta^2} + \frac{7}{9}\frac{m_a^2 - 16m_\pi^2/9}{m_a^2-m_{\eta'}^2} \right) \\ + \frac{5}{3}c_1 + c_2 + 2\sum_{\ell=e,\mu,\tau} c_\ell B(4m_\ell^2/m_a^2)\,,
    \end{multline}
    where $B(x) = 1 - x f(x)^2$ and
    \begin{equation}
        f(x) = \begin{cases}
        \text{sin}^{-1}\left(\frac{1}{\sqrt{x}}\right) & \text{ if } x \geq 1\\
        \frac{\pi}{2} + \frac{i}{2}\text{log}\left(\frac{1+\sqrt{1-x}}{1-\sqrt{1-x}}\right) & \text{ if } x < 1
            \,.
        \end{cases}
    \end{equation}
    The numerical factor of $-1.93$ arises from a numerical fit of a chiral perturbation theory model to decay and scattering data, with additional input from lattice QCD~\cite{GrillidiCortona:2015jxo}.
    
    \item \textit{Hadronic decay: 3$\pi$} 
    
    Because QCD is non-perturbative at the axion mass scale to which the SBN Program has sensitivity, modeling axion hadronic decays is quite complicated. The present work takes analytic results from \cite{AxionWidths}, as do all the phenomenological studies cited in this work.
    Note that the computation of these decay widths likely has large uncertainties. 

    There are two possible $3\pi$ decays: $\pi^+\pi^-\pi^0$ and $\pi^0\pi^0\pi^0$. The decay width for both is
    \begin{equation}
        \Gamma_{a\to 3\pi} = \frac{k}{2 S m_a}\int | A(a\to 3\pi)|^2 d\Phi_3
    \end{equation}
    where $S$ is a symmetry factor ($=1$ for $\pi^+\pi^-\pi^0$ and $=6$ for $\pi^0\pi^0\pi^0$), $k = 2.7$ is an empirical factor to fix the equivalent analytic result for the $\eta' \to 3\pi$ decay width to its experimental value, $A$ is the amplitude of the decay, and $d\Phi_3$ is the phase space integral. For a three-body decay of an ALP $a$ with mass $m_a$ into particles with masses $m_1$, $m_2$, and $m_3$, $\int d\Phi_3 = \int \frac{dm_{12}^2 dm_{23}^2}{128 \pi^3 m_a^2}$. The amplitudes are given by
    \begin{equation}
        \begin{split}
            A(a\to 3\pi^0) = \frac{m_\pi^2}{f_a f_\pi} &\left[ \langle a\pi \rangle - \delta_I \left(\frac{1}{\sqrt{3}} + \sqrt{2} S_{\eta\pi} + S_{\eta'\pi}\right)\left(\sqrt{2}\langle a \eta \rangle + \langle a \eta' \rangle\right) \right. \\
            & \left. + \sqrt{3}\delta_I \frac{m_\pi^2-2m_\eta^2}{m_\pi^2 - 4m_\eta^2}\left(\sqrt{2}S_{\eta\pi} + S_{\eta'\pi}\right) \right], \\
            A(a\to\pi^+\pi^-\pi^0) = \frac{1}{3 f_a f_\pi} & \left[\left(3m^2_{\pi^+\pi^-} - m_a^2  - 2 m_\pi^2\right) \langle a \pi \rangle \phantom{\frac{m_\eta^2}{m_\eta^2}}\right.\\ &\left.- \delta_I m_\pi^2 \left(\frac{1}{\sqrt{3}} + \sqrt{2} S_{\eta\pi} + S_{\eta'\pi}\right)\left(\sqrt{2}\langle a\eta \rangle + \langle a \eta' \rangle\right)\right.\\
            &\left. + \delta_I \frac{m_\pi^2 - 2m_\eta^2}{m_\pi^2-4m_\eta^2}\left(\sqrt{3}m_\pi^2 \left(\sqrt{2} S_{\eta\pi} + S_{\eta'\pi}\right) - 3 m^2_{\pi^+\pi^-} + m_a^2 + 3m_\pi^2\right) \right]\,,
        \end{split}
    \end{equation}
    where $\delta_I = (m_u - m_d) / (m_u + m_d) \approx 1/3$ is the isospin symmetry breaking factor, $S_{P P'} = M_{P P'} / (m_P^2 - m_{P'}^2)$ where $m_P$ is the mass of meson $P$, $M_{\pi\eta}^2 = -m_{\pi}^2\sqrt{2/3}$, $M_{\pi\eta'}^2 = -m_{\pi}^2/\sqrt{3}$, and $\langle \rangle$ denotes the expectation value of the enclosed states. These are
    \begin{equation}
        \begin{split}
            \langle a \pi \rangle &\approx \frac{\delta_I}{2}\frac{m_a^2}{m_a^2-m_{\pi}^2}, \\
            \langle a\eta \rangle &\approx \left[\frac{m_a^2}{\sqrt{6}} - \frac{m_{\pi}^2}{2\sqrt{6}}\right]\frac{1}{m_a^2 - m_\eta^2}, \\
            \langle a \eta' \rangle &\approx \left[\frac{m_a^2}{2\sqrt{3}} - \frac{m_{\pi}^2}{\sqrt{3}}\right]\frac{1}{m_a^2 - m_{\eta'}^2}\, .
        \end{split}
    \end{equation}

    \item \textit{Hadronic decay: $\pi^+\pi^-\gamma$} 
    
    This decay proceeds through the wide $\rho$ vector resonance $a\to\gamma(\rho\to\pi^+\pi^-)$. The decay width is given by
    \begin{equation}
        \Gamma_{a\to\gamma(\rho\to\pi^+\pi^-)} =\frac{3 \alpha m_a^3 }{2^{11} \pi^6f_a^2}\int dm_{\pi\pi}^2 |g^2 \text{BW}_\rho (m_{\pi\pi}^2) \langle a \rho\rho\rangle F(m_a)|^2 m^2_{\pi\pi} \left[1 - \frac{m_{\pi\pi}^2}{m_a^2}\right]^3 \left[1 - \frac{4m_\pi^2}{m_{\pi\pi}^2}\right]^{3/2}\, ,
    \end{equation}
    where $g \approx \sqrt{12\pi}$, $\text{BW}_\rho$ is the Breit-Wigner function for the $\rho$ resonance, and $F$ is a form factor that is fit from experimental data to be $1$ for $m_a <$ \SI{1.4}{GeV} (where the SBN Program is sensitive). The expectation value is given by
    \begin{equation}
        \langle a \rho\rho \rangle = \frac{\langle a \eta \rangle}{\sqrt{6}} + \frac{\langle a \eta' \rangle}{\sqrt{12}}\, .
    \end{equation}

    The Breitt-Wigner function has a non trivial dependence on the mass because the $\rho$ resonance is very wide. It is given by \cite{BaBarBW}
    \begin{equation}
        \begin{split}
            \text{BW}_\rho(s) &= \text{BW}^\text{GS}(s, m_\rho, \Gamma_\rho), \\
            \text{BW}^\text{GS}(s, m_\rho, \Gamma_\rho) &= \frac{1 + d(m_\rho) \Gamma_\rho/m_\rho}{m^2_\rho - s + f(s,m_\rho,\Gamma_\rho) - i m_\rho \Gamma(s,m_\rho,\Gamma_\rho)}, \\
            \Gamma(s,m_\rho,\Gamma_\rho) &= \Gamma_\rho \frac{s}{m^2_\rho}\left(\frac{\beta_\pi(s)}{\beta_\pi(m^2_\rho)}\right)^3, \\
            \beta_\pi(s) &=\sqrt{1 - 4m_\pi^2/s}, \\
            d(m_\rho) &= \frac{3}{\pi}\frac{m_\pi^2}{k^2(m^2_\rho)}\text{ln}\left(\frac{m_\rho + 2 k(m^2_\rho)}{2m_\pi}\right) +\frac{m_\rho}{2\pi k(m^2_\rho)}-\frac{m_\pi^2 m_\rho}{\pi k^3(m^2_\rho)}, \\
            f(s,m_\rho,\Gamma_\rho) &= \frac{\Gamma_\rho m^2_\rho}{k^3(m^2_\rho)} \left[k^2(s)(h(s) -h(m^2_\rho)) + (m^2_\rho-s)k^2(m^2_\rho)h'(m^2_\rho) \right], \\
            k(s) &= \frac{1}{2}\sqrt{s}\beta_\pi(s), \\
            h(s) &= \frac{2}{\pi} \frac{k(s)}{\sqrt{s}}\text{ln}\left(\frac{\sqrt{s} + 2k(s)}{2m_\pi}\right), \\
            h'(s) &= \frac{dh(s)}{ds}\, .
        \end{split}
    \end{equation}
    
    \item \textit{Other hadronic decays:} 
    
    Other decays to hadronic final states such as $4\pi$, $\eta\pi\pi$, etc.\:are not significant at the masses that the SBN Program is sensitive to ($m_a <$ \SI{900}{MeV}) so they are not implemented in \lstinline{MeVPrtl}.
    
    \item \textit{Leptonic decays:} 
    
    The decay width to a lepton with tree-level coupling presented by \cref{eqn:axionLepton} is given by
    \begin{equation}
        \Gamma_{a\to\ell\ell} = \frac{c_\ell^2m_a m_\ell^2}{8\pi f_a^2}\sqrt{1 - \frac{4m_\ell^2}{m_a^2}}.
    \end{equation}

    Fortuitously, leptonic decays can also arise from the other gauge couplings at loop-level, qualifying leptonic decays as a viable search channel even in minimalist scenarios that do not explicitly include the leptonic Lagrangian term (see \cite{Berger_Putnam_2024} and references therein). Under such a scenario, the induced leptonic decays depend on the relative strengths of the gauge couplings $c_1$, $c_2$, and $c_3$.  To describe this scenario, \cite{Berger_Putnam_2024} assumed values of $c_1,c_2,c_3$ defined at an energy scale of $4\pi f_a$.  Starting with the initial condition that $c_\ell = 0$ at this scale, they then solve the one-loop renormalization group equations to run the couplings down to the scale of the top mass $m_t \approx 172.69~{\rm GeV}$~\cite{PDG2020}.  A non-zero $c_\ell$ is generated by this evolution. Below the top mass, the running of $c_\ell$ stops at one loop, so it is considered to be the low-scale effective value of this coupling. 

    For all intents and purposes, $c_\ell = 0$ is configured as a free parameter in \lstinline{MeVPrtl}. The choice of $c_\ell$ is left to the analyzer, and events can always be reweighted to a different value of $c_\ell$ after generation.

\end{enumerate}

The photonic, leptonic, $3\pi$, and $2\pi\gamma$ hadronic decays are all significant at various values of the axion mass, so all of them are included when determining the axion lifetime, i.e.\:the probability that the axion will reach the detector. However, only decays to the photonic and leptonic final states are included in \lstinline{MeVPrtl}. Adding the hadronic decays would require simulation of the angular dependence of those decays.

\section{Discussion and Outlook}
\label{sec:discussion}

The \lstinline{MeVPrtl} generator, as described here, provides implementations of the HPS, HNL, and ALP BSM models for the SBN Program Monte Carlo (MC) simulation. The generated events include timing information, enabling valuable studies given LArTPC experiments' recently demonstrated ability to distinguish events with nanosecond precision (see Ref.~\cite{MicroBooNE:2023ldj} for an overview). The generator is implemented in a modular way that is designed to consolidate shared phenomenology between models and be extensible to other model implementations of further interest. Future extensions could include alternate axion models such as leptophilic axion-like particles \cite{Bertuzzo:2022fcm}, vector portal models such as dark photons \cite{DeRomeri:2019kic}, or inelastic dark matter that couples through a vector portal, among others. Dark bremsstrahlung simulation would need to be added to \lstinline{MeVPrtl} for the production of vector portal particles. Additionally, any models that involve scattering in the detector volume would also require modifications to the generator, beyond simply modifying the relevant decay widths.

While \lstinline{MeVPrtl} already serves as a valuable tool for analyses in the SBN Program, there are some limitations that the user should be aware of. On the theory side, the determination of production of these particles can be challenging. Semi-hadronic kaon decays rely on chiral perturbation theory and the uncertainties inherent therein. More importantly, there is as yet no consistent way to determine ALP production through mixing with neutral mesons. The mixing angle used in this calculation is basis-dependent. The kinematics are only fully consistent in the infinite momentum limit. Rectifying this issue is fraught with theory challenges and is beyond the scope of this work. Away from the ALP mass matching a neutral meson mass and for boosted ALP production, a reasonable approximation is obtained.

In terms of the decays of the portal particles, the hadronic channels are subject to the same theoretical caveats in general as for production. For the ALP case in particular, there are several different determinations of $c_\gamma$ under different assumptions about the flavors involved and the handling of axion-meson mixing (see for example \cref{eqn:cgamma}).

The production of the portal particles in \lstinline{MeVPrtl} also inherits uncertainties from the modeling of their parent particles' flux. The particles produced in kaon decays inherit uncertainties from the beamline modeling, typically performed with tools such as \lstinline{g4bnb}~\cite{MiniBooNE:2008hfu,MicroBooNE:2018efi,Paton:2025lmv} and \lstinline{g4numi}~\cite{Wood:2024jos}, which rely on \lstinline{Geant4}~\cite{Asai:2006qm,Allison:2006ve,Asai:2015xno}. Driven by extensive effort to resolve neutrino oscillation anomalies, there already has been, and continues to be, extensive effort to quantify and suppress the flux uncertainties relevant for neutrino processes (see, e.g. \cite{TonyWoodThesis}), some of which also apply for BSM processes. However, there is some uncertainty associated with how to extrapolate these tools to BSM processes with different kinematics. Further, the production of neutral mesons relies on \lstinline{Pythia8}, which does not include secondary interactions and has its own set of uncertainties in terms of soft QCD simulation.  Finally, the calculations of certain decays such as $a \to 3\pi$ face large uncertainties.

Other codes have been developed to simulate portal particles at fixed-target experiments more broadly.  In particular, there has been some focus on HNLs, such as in the code \lstinline{BeamHNL} \cite{BeamHNL_code}, a \lstinline{GENIE}-based code that models HNL decays similar to the ones described above.  \lstinline{MeVPrtl} further considers the other models described here, and is extensible to even more models. There are also codes like \lstinline{MadDump} \cite{MadDump_code} and \lstinline{BdNMC} \cite{BdNMC_code}, which can model some of the same physics over some of the relevant parameter space, but do not include a full modeling of the beamline and focusing. Of these, \lstinline{MeVPrtl} is the only generator fully integrated into the \lstinline{LArSoft}-based SBN Program simulation pipeline.

The \lstinline{MeVPrtl} event generator was used for signal event generation for the ICARUS Collaboration's first published physics result, which selected reconstructed pairs of fiducial muon tracks to probe hidden scalars emerging from kaon decay \cite{icarus2024hiddensector}. \lstinline{MeVPrtl} is anticipated for use in forthcoming BSM searches led by ICARUS and SBND as well. Given its efficient organization and full integration into the \lstinline{LArSoft} simulation and analysis pipeline, \lstinline{MeVPrtl} is an exemplary BSM generator for other LArTPC-based experiments such as MicroBooNE and DUNE. If ported outside of the LArSoft framework, it could further serve as a valuable tool for other experiments such as MINER$\nu$A and NO$\nu$A.

\section*{Acknowledgments}

This document was prepared by the ICARUS and SBND Collaborations using the resources of the Fermi National Accelerator Laboratory (Fermilab), a U.S. Department of Energy, Office of Science, HEP User Facility. Fermilab is managed by FermiForward Discovery Group, LLC, acting under Contract No. 89243024CSC000002.

This work was also supported by Istituto Nazionale di Fisica Nucleare (INFN, Italy), EU Horizon 2020 Research and Innovation Program under the Marie Sklodowska-Curie Grant Agreement Nos. 822185, 858199, and101003460 and Horizon Europe Program research and innovation programme under the Marie Sklodowska-Curie Grant Agreement No. 101081478, the research contract per Law 240/2010, Art. 24 (3)(a), and D.G.R. 693/2023 (REF. PA:2023-20090/RER—CUP:J19J23000730002) by FSE+ 2021–2027. Furthermore the support of CERN in the detector overhauling within the Neutrino Platform framework, in the detector installation and commissioning, is acknowledged. This work was also supported by the Anusandhan National Research Foundation (ANRF, India) under the Ramanujan Fellowship (Grant No. RJF/2025/000203).

The ICARUS Collaboration would like to thank the MINOS Collaboration for having provided the Side CRT panels as well as Double Chooz (University of Chicago) for the Bottom CRT panels. We also acknowledge the contribution of many SBND colleagues, in particular for the development of a number of simulation, reconstruction and analysis tools which are shared within the SBN program.

The SBND Collaboration acknowledges the generous support of the following organizations: the U.S. Department of Energy, Office of Science, Office of High Energy Physics; the U.S. National Science Foundation; Los Alamos National Laboratory for LDRD funding, the Science and Technology Facilities Council (STFC), part of United Kingdom Research and Innovation (UKRI), the UKRI Future Leaders Fellowship (grant number MR/V022407/1), and The Royal Society; the Swiss National Science Foundation; the Spanish Ministerio de Ciencia, Innovacíon y Universidades (MICIU/ AEI/ 10.13039/ 501100011033) under grants No PRE2019-090468, CNS2022-136022, RYC2022-036471-I, PID2023-147949NB-C51 \& C53 and Comunidad
de Madrid (PEJ-2023-AI/COM-28399); the European Union’s Horizon 2020 research and innovation program under GA no 101004761 and the Marie Sklodowska-Curie grant agreements No 822185, 101081478, and 101003460; the São Paulo Research Foundation 1098 (FAPESP), the National Council of Scientific and Technological Development (CNPq) and Ministry of Science, Technology \& Innovations-MCTI of Brazil; the Minas Gerais research foundation (FAPEMIG), grants APQ-00544-23 and APQ-01249-24; the Anusandhan National Research Foundation (ANRF, India) under the Ramanujan Fellowship (Grant No. RJF/2025/000203). An award of computer time was provided by the ASCR Leadership Computing Challenge (ALCC) program. This research used resources of the Argonne Leadership Computing Facility, which is a U.S. Department of Energy Office of Science User Facility operated under contract DE-AC02-06CH11357.

\bibliographystyle{elsarticle-num}
\bibliography{mevprtl}

\pagebreak
\appendix

\section{Generating HPS Events with \lstinline{MeVPrtl}}
\label{apn:runHPS}
For the HPS model, there are several settings in \lstinline{icaruscode} that allow for event generation. Note that \lstinline{icaruscode} depends on \lstinline{sbncode}.  Both are part of the broader SBNSoftware GitHub organization.  The \linebreak\lstinline{dissonant_higgs_gen.fcl} will generate an art-ROOT file with HPS decays. The \lstinline{dissonant_higgs_tree_icarus.fcl} will generate a ROOT tree of HPS decays.  The \lstinline{dissonant_higgs_corsika_p_gen} file will overlay \lstinline{CORSIKA}-generated cosmic rays on the HPS events.  By default, $\theta = 10^{-5}$. All kinematically allowed HPS decays are enabled by default.
To set all the parameters, one may use a file such as
\begin{lstlisting}[language=C]
#include dissonant_higgs_gen.fcl

higgsM: 0.3 # GeV / c^2
MixingAngle: 1e-5

#include set_higgsM.fcl
physics.producers.generator.Flux.MixingAngle: @local::MixingAngle
physics.producers.generator.Decay.ReferenceHiggsMixing: @local::MixingAngle
\end{lstlisting}

where \lstinline{higgsM} sets the HPS mass $m_S$ and \lstinline{MixingAngle} sets $\theta$.

\pagebreak
\section{Generating HNL Events with \lstinline{MeVPrtl}}
\label{apn:runHNL}
The primary file for HNL production in \lstinline{icaruscode} is via the configuration file \lstinline{hnl_icarus_gen.fcl}, which will generate an art-ROOT file with HNL decays. By default, $|U_{e4}|^2 = 0$, $|U_{\mu 4}|^2 = 10^{-7}$, and $|U_{\tau 4}|^2 = 10^{-5}$ (for $\tau$ decay production).  The default production is via kaon decay.  The default decay of the HNL is to $\mu \pi$.  The HNL is taken to be a Dirac fermion by default.  To set all the parameters, one may use a file such as, for example, 

\begin{lstlisting}[language=C]
#include hnl_icarus_gen.fcl
hnlM: 0.3
MagUe4: 0
MagUm4: 1e-5
Majorana: true
Decays: ["nu_mu_mu"]

#include set_hnlM.fcl
physics.producers.generator.Flux.MagUe4: @local::MagUe4
physics.producers.generator.Decay.ReferenceUE4: @local::MagUe4
physics.producers.generator.Flux.MagUm4: @local::MagUm4
physics.producers.generator.Decay.ReferenceUM4: @local::MagUm4
physics.producers.generator.Decay.Majorana: @local::Majorana
physics.producers.generator.Decay.Decays: @local::Decays
\end{lstlisting}

where \lstinline{hnlM} sets the HNL mass $m_N$ in GeV/c$^2$, \lstinline{MagUe4}, \lstinline{MagUm4} set $|U_{e4}|^2$, $|U_{\mu 4}|^2$ respectively, \lstinline{Majorana} sets whether the HNL is Dirac or Majorana, and \lstinline{Decays} sets the decay modes of the HNL. 

The available decay modes for the HNL are:
\begin{itemize}
\item \lstinline{"mu_pi"}: $\mu\pi$
\item \lstinline{"e_pi"}: $e\pi$
\item \lstinline{"nu_mu_mu"}: $\nu\mu^+\mu^-$
\item \lstinline{"nu_e_e"}: $\nu e^+e^-$
\item \lstinline{"nu_pi0"}: $\nu\pi^0$
\item \lstinline{"nu_eta"}: $\nu\eta$
\item \lstinline{"nu_etap"}: $\nu\eta^\prime$
\end{itemize}

\pagebreak
\section{Generating Heavy QCD Axion Events with \lstinline{MeVPrtl}}
\label{apn:runALP}
There are several configurations of ALP generation in \lstinline{MeVPrtl} available in \lstinline{icaruscode}.  The \lstinline{alp_tree_icarus.fcl} configuration will generate a ROOT \lstinline{TTree} of axion decays. The \lstinline{alp_gen_icarus.fcl} will generate an art-ROOT file with axion decays. The \lstinline{alp_corsika_p_gen_icarus.fcl} configuration generates axion decays with \lstinline{CORSIKA} cosmic muons.  Event generation can be run by using a \lstinline{fihcl} file including one of these files and setting the desired parameters.  By default, $c_\ell = c_\mu =0.01$ and $c_1 = c_2 = c_3 = 1$.  To set all the parameters, one may use a file such as, for example, 
\begin{lstlisting}[language=C]
#include alp_gen_icarus.fcl

alpM: 0.3 # GeV /c^2
fa: 1e5   # GeV
cAl: 0.01
cB: 1
cW: 1
cG: 1

#include set_alpM.fcl
#include set_alpfa.fcl
physics.producers.generator.Flux.cAl: @local::cAl
physics.producers.generator.Decay.ReferenceALPcAl: @local::cAl
physics.producers.generator.Flux.cB: @local::cB
physics.producers.generator.Decay.ReferenceALPcB: @local::cB
physics.producers.generator.Flux.cW: @local::cW
physics.producers.generator.Decay.ReferenceALPcW: @local::cW
physics.producers.generator.Flux.cG: @local::cG
physics.producers.generator.Decay.ReferenceALPcG: @local::cG
\end{lstlisting}

where \lstinline{alpM} sets the ALP mass $m_a$, \lstinline{fa} sets the decay constant $f_a$, \lstinline{cAl} sets the axial lepton coupling $c_\ell$, and \lstinline{cB}, \lstinline{cW}, \lstinline{cG} set $c_1$, $c_2$, $c_3$ respectively.

\end{document}